\documentstyle[aaspp,12pt]{article}

\def\kp {{\bf K}_\perp}
\def\pt{polarizability tensor}
\def\Mesz{M\'esz\'aros}
\def \al{\alpha}
\def \om{\omega}
\def \de{\delta}

\begin{document}

\newcommand{\be}{\begin{equation}}
\newcommand{\ee}{\end{equation}}
\newcommand{\lapr}{\raisebox{-.6ex}{\mbox{
$\stackrel{<}{\mbox{\scriptsize $\sim$}}\:$}}}
\newcommand{\gapr}{\raisebox{-.6ex}{\mbox{
$\stackrel{>}{\mbox{\scriptsize $\sim$}}\:$}}}
\title{POLARIZATION MODES IN
A STRONGLY
MAGNETIZED HYDROGEN GAS}

\author{Tomasz Bulik}
\affil{ University of Chicago, Astr.\& Astrophysics and Enrico Fermi Institute,
 5640 S. Ellis Av., Chicago, IL 60615}
\and
\author{George G.~Pavlov
\footnote{On leave from A.~F.~Ioffe Physico-Technical Institute,
St.~Petersburg, 194021, Russia}}
\affil{
 Pennsylvania State University, 525 Davey Lab., University Park, PA 16802}
\centerline{E-mail: tomek@gamma.uchicago.edu, pavlov@astro.psu.edu}
\begin{abstract}
Propagation of high-frequency radiation in an anisotropic medium
can be described in terms of
 two normal modes with
 different polarizations and different absorption
coefficients.
We investigate the properties of the normal modes
in a strongly magnetized hydrogen gas for conditions
expected in atmospheres of
isolated neutron stars.
We use the  Kramers-Kronig  relations to obtain the \pt\
for the strongly magnetized hydrogen.
We derive and compute the polarizations and absorption
coefficients of the normal modes from the \pt\
using both analytical approximations and
 numerical calculations.  We find that
the spectral features and anisotropy associated with the bound-bound
and bound-free transitions in the magnetized hydrogen
are manifested in
the polarization characteristics,
which affects substantially the
spectral and angular dependences of  the absorption coefficients
of the normal modes and the transfer of radiation in neutron
star atmospheres.
There exist at least two critical frequencies
where, for any direction of propagation,
either the normal mode polarizations are exactly linear
or orientations of two polarization ellipses coincide with each other.
For each of these frequencies, there exist a direction of propagation
for which the two normal modes are linearly polarized,
and their polarizations and absorption coefficients fully coincide.
The unusual properties of the normal modes should
manifest themselves in the spectra, angular distribution,
and polarization of the thermal-like radiation emitted from
surface layers of neutron stars.
\end{abstract}


\section{ INTRODUCTION}
With the development of observational capabilities of space X-ray/UV
observatories and ground based telescopes it is now feasible to detect
radiation from surface layers of neutron stars with rather low temperatures,
$T\sim (0.2-10)\times 10^5$ K (\"Ogelman 1995; Pavlov, Stringfellow
\& C\'ordova 1995). The immense magnetic fields
of  neutron stars, $B\sim 10^{11} - 10^{13}$ G,
substantially increase the ionization potentials of atoms (e.~g.,
Canuto \& Ventura 1977):
$I\sim Z^2 {\rm Ry}\ln^2(\gamma /Z^2)$ at
$\gamma\gg Z^2$, where $\gamma =\hbar\omega_{Be}/(2{\rm Ry})$=
$B/(2.35\times
10^9$ G) is the magnetic field in the atomic units, $\omega_{Be}=eB/m_ec$ is
the
electron cyclotron frequency, $Z$ is the effective ion charge, and
${\rm Ry}=m_ee^4/2\hbar^2=13.6$ eV is the Rydberg energy. This means that
even light atoms may be only partly ionized (or even fully nonionized)
in   neutron star atmospheres,
and be the main contributors to the atmosphere opacity (Pavlov et al.~1995a).
Thus, to model the atmospheres for further interpretation of the observational
results, it is necessary to investigate the  effect of the strongly magnetized
atoms
on absorption and propagation of radiation.

It is well known (see, e.g., Gnedin \& Pavlov 1974) that radiation propagates
in an anisotropic medium in form of two so-called normal modes
(NMs)  which have
different polarizations and, consequently, different
absorption and refraction coefficients.
The dependence of the NM absorption coefficients on the frequency
and direction of propagation determines the spectrum and angular
distribution of radiation emitted from an anisotropic medium.
Properties of the NMs have been thoroughly investigated
(e.g., Pavlov, Shibanov \& Yakovlev 1980;
Kaminker, Pavlov \& Shibanov 1982,
1983; M\'esz\'aros 1992, and references therein) for the case of fully
ionized plasma, when polarization and absorption of the modes are
determined
by the free-free transitions and Thomson scattering in the magnetic field.
However, there has been no analysis of how these properties would change if
the bound-bound and bound-free transitions in the strongly magnetized atoms
are important.

To find the polarizations of the normal modes, one should proceed from
the polarizability tensor of the anisotropic medium. The contribution of the
bound-bound and bound-free transitions to the components of
the tensor is determined by the structure of the strongly magnetized
atoms. For the simplest hydrogen atom, assuming the center of mass
of the atom is fixed,
the atomic structure and probabilities
of the radiative transitions, which determine the antihermitian part
of the polarizability tensor, have been investigated in detail (e.~g.,
Canuto \& Ventura 1977, Forster et al.~1984, Potekhin \& Pavlov 1993).
The hermitian part of the polarizability has not been discussed, to the
best of our knowledge, even under this oversimplified assumption.
Moreover, it has been
understood recently that motion of  atoms in strong magnetic fields
changes qualitatively their structure (Herold et al.~1981; Vincke \& Baye
1988; Vincke, Le Dourneuf \& Baye 1992;
Pavlov \& M\'esz\'aros 1993; Potekhin 1994).
This occurs because the motion across the magnetic field induces an
electric field, ${\bf F} =c^{-1}{\bf V}\times {\bf B}$, which breaks the
cylindrical symmetry and couples the center-of-mass motion to the
internal structure of the atom.  The energies and wave functions of the
moving atom depend on the transverse component $\kp$
 of a generalized momentum
${\bf K}$. When $\kp$ is small (at low  temperatures of the gas),
the perturbation approach is applicable,
and the coupling effect can be described in terms of an effective
transverse mass of the atom (Vincke \& Baye 1988; Pavlov \& M\'esz\'aros
1993). This transverse mass always exceeds the mass of the atom without the
magnetic field. It grows with the magnetic field, being and  is  generally
higher for more excited atomic states. When $\kp$ increases beyond
the applicability of the perturbation theory, the induced electric
field causes decentering of the atom, i.~e., the electron gets into a
potential well which is shifted from the proton   more for the larger
$\kp$. These decentered atoms have very unusual properties. For instance,
their velocity decreases with increasing $\kp$ because the growth of
$\kp$ goes not to acceleration of the atom but to larger decentering.
The decentered states have relatively small binding energies, so
that they can be occupied only at relatively high temperatures and not too
high densities, when the size of the atom does not exceed the  mean
distance between the particles.

Qualitative effects of the coupling on the radiative transitions were
discussed by Pavlov \& \Mesz~(1993). They show, in particular, that
different dependence of different atomic levels on $\kp$ leads to
a new broadening mechanism of the spectral lines and photoionization
edges. The ``magnetic coupling width'' exceeds the Doppler width by orders
of magnitude, growing proportional to the gas temperature.
 More detailed treatment of the radiative transitions, with
allowance for the decentering of moving atoms, has been presented
by Bezchastnov \& Potekhin (1994) for   bound-free transitions
and by Pavlov \& Potekhin (1995) for   bound-bound transitions.

In the present paper we will consider the \pt\ and properties of the
polarization modes of the hydrogen gas for the case of relatively low
temperatures, when interaction of the radiation with the medium is
mainly determined by the nonionized atoms. At these low temperatures,
most of the atoms occupy the ground level, and, at densities typical
for   neutron star atmospheres, one can neglect contribution of
the decentered states. In these conditions, the coupling effects
can be considered in the frame of the perturbation approach.
In Section 2 we recall general equations for the polarizability
tensor, normal mode polarizations and absorption coefficients,
and describe our approach for calculation of these quantities.
We present numerical results and analytical
limits   in Section 3.
In Section 4 we discuss some consequences of the obtained results  for
neutron star atmosphere modeling.

\section{ BASIC EQUATIONS}

\subsection{Polarization Characteristics of the  Normal Modes.}

The propagation of  electromagnetic waves in a magnetized
medium is fully described by the dielectric tensor
\begin{equation}
\epsilon_{ik}= \delta_{ik}+4\pi\chi_{ik}~,
\end{equation}
where $\chi_{ik}$ is the polarizability tensor.
In the case of a tenuous medium, (i.~e., at $|n_j - 1|< 1$, where $j=1,2$
labels
the NM, and $n_j$ is the complex refraction index) the NMs
are transverse, and the (real) refraction index $\kappa_j$ and
the absorption
coefficient $\mu_j$ are
given by (Pavlov et al.~1980)
\begin{equation}
n_j=\kappa_j+i\frac{c}{2\om}\mu_j= n_I\pm \sqrt{ n_L^2 + n_C^2}~,
\end{equation}
where
\begin{equation}
n_I = 1+\pi(\chi_{yy} + \chi_{xx}\cos^2\theta-\chi_{xz}\sin 2\theta
+\chi_{zz}\sin^2\theta )~, 
\ee
\be
n_L = \pi(\chi_{yy}-\chi_{xx}\cos^2\theta+\chi_{xz}\sin 2\theta
-\chi_{zz}\sin^2\theta)~, 
\ee
\be
n_C = 2\pi i (\chi_{xy}\cos\theta + \chi_{yz}\sin\theta )~.
  \label{ens}
\end{equation}
Here $\theta$ is the angle between the wave vector ${\bf q}$ and
the magnetic field $\bf B$, and $\chi_{ik}$ is the polarizability
tensor in the coordinate frame with the $z$-axis along $\bf B$
 and the $x$-axis in the $\bf B$-$\bf q$ plane.
The NMs can be described by
the ellipticity ${\cal P}_j$ (the modulus $|{\cal P}_j|$
is the ratio
of the minor axis to the major axis of the polarization ellipse;
the sign of ${\cal P}_j$ determines the direction of rotation
of the polarization vector)
 and the position angle $\delta_j$
between the major axis of the polarization  ellipse and the
projection of $\bf B$ onto the plane perpendicular to $\bf q$.
The ellipticity  and position angle  are determined by the modulus
$r_j$ and phase $\varphi_j$ of the complex quantity
\begin{equation}
r_j \exp({i\varphi_j})
= {-n_C\pm\sqrt{n_L^2+n_C^2}\over n_L} ~;
\end{equation}
\be
{\cal P}_j = {r_j-1\over r_j+1},\qquad
\delta_j = {\varphi_j\over 2}~.
\ee
It follows from these equations that the NM properties
can be conveniently analyzed with the
use of real parameters $q$ and $p$ defined as (Gnedin \& Pavlov 1974)
\begin{equation}
q+ip={n_L\over n_C}~.  \label{qpdef}
\end{equation}
 The NM polarizations are nearly orthogonal ($\delta_2\simeq
\delta_1\pm\pi/2$; $\delta_j\simeq 0$ or $\pi/2$;
and ${\cal P}_2= -{\cal P}_1$) when
$2 |p| \ll p^2+q^2+1$. In particular,
they are orthogonal when
$p^2\ll q^2$ (the most common case, which is usually realized
in a wide range  of frequencies and angles),
or when $p^2\gg q^2$ and $|p|\gg 1$. In the case
when $p^2\ll q^2$, the (orthogonal) NMs are linearly polarized,
$|{\cal P}_1|=|{\cal P}_2| \simeq 0$, if
$|q|\gg 1$, and they are  circularly polarized,
$|{\cal P}_1|= |{\cal P}_2|\simeq 1$, if $|q|\ll 1$.
If $p^2 \gg q^2$ and $q^2\ll 1$, then the NMs are
linearly polarized
at $|p|\geq 1$,
and the polarization ellipses coincide [$\delta_1\simeq \delta_2
\simeq (\pi/4)\, {\rm sign}(p)$, ${\cal P}_1=-{\cal P}_2$] at $|p|\leq 1$.
In the special case  when $q=0$ and $|p|=1$, the NM polarizations
are purely linear and coincide with each other
(are ``completely non-orthogonal''): ${\cal P}_1
={\cal P}_2 = 0$, $\delta_1=\delta_2=(\pi/4)\, {\rm sign}(p)$.
 For further details see, e.~g., Pavlov et al.~(1980).

\subsection{General Equations for the Polarizability
Tensor in the Dipole Approximation.}

 In the dipole approximation, which is justified for
 frequencies that are not too high (see, e.~g., Pavlov \& Potekhin 1995
for a discussion), the polarizability tensor
of a magnetized gas is diagonal
in the cyclic coordinates,
$\chi_{\alpha\beta} = \chi_\alpha \delta_{\alpha\beta}$,
where $\alpha, \beta = 0, \pm 1$, and the cyclic unit
vectors are defined as
${\bf e}_0=e_z$, ${\bf e}_{\pm 1}=2^{-1/2} ({\bf e}_x \pm i{\bf e_y})$.
 In terms of
the cyclic components $\chi_\alpha$, we have
\begin{equation}
 n_I = 1 + \pi\left[ \left(\chi_{+1}+\chi_{-1}\right){1+\cos^2\theta\over
2} + \chi_0 \sin^2\theta \right]~, 
\ee
\be
n_L = \pi\left({\chi_{+1}+\chi_{-1}\over 2} -\chi_0\right)\sin^2\theta~,
\ee
\be
n_C = \pi \left(\chi_{-1}-\chi_{+1}\right)\cos\theta~,
\end{equation}
and
\begin{equation}
q+ip=(\tilde{q}+i\tilde{p})\, {\sin^2\theta\over 2\cos\theta}~,\qquad
\tilde{q}+i\tilde{p}={\chi_{+1}+\chi_{-1}-2\chi_0\over \chi_{-1}-\chi_{+1}}\ .
\end{equation}
The polarizability tensor is the sum
of the hermitian and antihermitian
parts, $\chi_\alpha =\chi_\alpha^H +i\chi_\alpha^A$,
responsible for refraction and absorption of radiation.
The antihermitian part
can be written
(for a single component ideal gas) as
\begin{eqnarray}
\chi_\alpha^A & = & {\omega_{pa}^2\over 8\omega}\sum_{if} \xi_i\
\left[ f_{if}^\alpha\ \delta (\om -\om_{fi}) - f_{fi}^\al\
 \de (\om
-\om_{if})\right] \\
& = &\frac{\omega_{pa}^2}{8\omega}{\sum_{if}}' (\xi_i-\xi_f)
\left[f_{if}^\alpha\ \delta(\omega-\omega_{fi})-
f_{fi}^\alpha\ \delta(\omega-\omega_{if})\right]~,
\label{antiherm}
\end{eqnarray}
where $\omega_{pa}=(4\pi n_a e^2/m_e)^{1/2}$ is
the ``plasma frequency'' of   bound electrons, $n_a$ is the number
density of atoms, $\xi_i =n_{ai}/n_a$ is the population of the
quantum state $i$, $\omega_{fi}=(E_f-E_i)/\hbar$
is the transition frequency,
and $f_{if}$ is the oscillator strength for the transition
$i\to f$,
\begin{equation}
f_{if}^\al = {2m_e\om_{fi}\over \hbar} |({\bf r}_{fi}\cdot {\bf e}_\al )|^2
= {2m_e\om_{fi}\over\hbar} |(x_{-\al})_{fi}|^2~.
\label{ocsstrength}
\end{equation}
The double sum $\sum_{if}$ implies summation (and integration
over the continuum part of the energy spectrum) over
all possible combinations of $i$ and $f$ (except $i=f$); in the sum
$\sum_{if}'$ only the states are summed (integrated) for which
$E_f > E_i$.
The general equation for the hermitian part can be written,
in the same approximation, as
\begin{equation}
\chi_\al^H = -{\om_{pa}^2\over 8\pi\om} \sum_{if} \xi_i
\left[{f_{if}^\al \over \om -\om_{fi}} - {f_{fi}^\al \over \om -\om_{if}}
\right]
=-\frac{\omega_{pa}^2}{8\pi\omega}{\sum_{if}}'(\xi_i-\xi_f)
\left[\frac{f_{if}^\al}{\omega-\omega_{fi}} - \frac{f_{fi}^\al}
{\omega -\omega_{if}}\right]
{}~.  \label{hermitian}
\end{equation}
 It follows from the completeness of the wave
functions that
\begin{equation}
\sum_f {1\over \om_{fi}} \left(f_{if}^\al + f_{fi}^\al \right) = 0~,
\label{complet}
\end{equation}
which means, in particular, that $\chi_\al^H$ has no   pole at $\om =0$ if all
the
electrons are bound.

We see from these equations that direct calculation of $\chi^H$ implies an
additional integration over the continuum states,
as compared to $\chi^A$, because there are no
delta-functions in equation~(\ref{hermitian}), and
this integration may not be   straightforward. An alternative
way to find the hermitian part is to use the Kramers-Kronig (KK)
relation (Guilbert transformation of the antihermitian part):
\begin{equation}
\chi_\al^H(\om ) = {\cal{P}\over\pi}\int_{-\infty}^\infty
{\chi_\al^A(\om')\over \om'-\om }\ {\rm d}\om'\equiv \hat{K}[\chi_\al^A(\om
)]~,
\label{kram-general}
\end{equation}
which is derived from the general analytical properties of the polarizability:
$\chi(\omega)$ is a regular function in the upper half-plane of
complex $\omega$; $\chi(\omega)$ has no singularities on the real axis
except possibly a simple pole at the origin; $\chi(\omega\to\infty)\to
 0$  (see e.g. Landau \&\ Lifshitz, 1960).
Since the quantity $\om\chi_\alpha(\omega)$
also has all the necessary analytical
properties, we may apply the KK
 transformation to this product
\begin{equation}
\chi_\al^H(\om ) ={1\over\omega} \hat{K} [\om\chi_\al^A(\om )]
= {c\over 4\pi\omega} \hat{K}[\mu_\alpha(\omega)\, \Theta(\omega)+
\mu_{-\alpha}(-\omega)\,\Theta(-\omega)]~,
\label{kramkron}
\end{equation}
where $\Theta(x) =1$ for $x\ge 0$, $\Theta(x)=0$ for $x<0$,
$\mu_\alpha(\omega)$ is the absorption coefficient (defined
for $\omega >0)$:
\begin{equation}
\mu_\al (\om )=
{\pi\om_{pa}^2\over 2c}
{\sum_{if}}' (\xi_i -\xi_f)\ f_{if}^\alpha\ \delta(\omega-\omega_{fi})
=n_a [1-\exp(-\hbar\omega/kT)]\sum_i \xi_i \sigma_i(\omega)~,
\label{abscoef}
\end{equation}
where $\sigma_i(\omega)$ is the absorption cross section, and $T$
is the temperature. To derive equation (\ref{kramkron}), we used
the property of the oscillator strength
\be
f_{if}^\alpha = - f_{fi}^{-\alpha}
\label{fminf}
\ee
which implies the symmetry relations
\be
\chi_\al^A(-\omega) = -\chi_{-\al}^A(\omega), \qquad
\chi_\al^H(-\omega) = \chi_{-\al}^H(\omega)~.
\ee
Applying the KK transformation to $\om \chi_\al$ instead of $\chi_\al$
is technically easier because the additional $\om$ cancels the false
pole (for bound electrons) of $\chi_\al$ at $\om =0$. Thus, to find
$\chi^H$ in this way, one needs the absorption coefficients
at $\omega >0$
for the longitudinal ($\al =0$), right ($\al =+1$), and left
($\al =-1$) polarizations.

\subsection{Oscillator Strengths and Cross Sections  for
a Strongly Magnetized Hydrogen Atom
at Rest}

The absorption coefficients for the bound atoms are contributed from
the bound-bound and bound-free transitions. For isolated atoms at rest,
the problem of calculating the {\it bound-bound} absorption coefficients
is reduced to calculating the oscillator strengths $f_{fi}^\al$.
 When the magnetic field ${\bf B}=B{\bf e}_z$ is so strong
that the ratio of
the electron cyclotron energy, $\hbar\omega_{Be}=2\gamma{\rm Ry}
= 11.6 B_{12}$ keV,
substantially exceeds
the binding energy,
$\sim {\rm Ry}\,\ln^2\gamma$ at $\ln^2\gamma\gg 1$,
the adiabatic approximation is valid. In this approximation, the
bound states of the hydrogen atom
are described by the Landau number $N=0,1,2,...$, the negative of the
$z$-projection of the angular momentum $s=-N, -N+1, -N+2,....$,
and the longitudinal quantum number $n=0,1,2,...$ (see, e.~g., Canuto
and Ventura 1977 for details). The wave functions are the products of
the Landau states $\Phi_{Ns}({\bf r_\perp})$ and longitudinal wave
functions $g_{Nsn}(z)$ which obey a one-dimensional Schr\"odinger
equation. The oscillator strengths of the transitions between the
states $\kappa = (N s n)$ and $\kappa'=(N' s' n')$ are given by
the following equations
\begin{eqnarray}
f_{\kappa\kappa'}^{-1} & =& {2\omega_{\kappa'\kappa}\over\omega_{Be}}
[(N+s)\delta_{N'N}+(N+1)\delta_{N',N+1}]\ |Z_{\kappa'\kappa}^{(0)}|^2
\delta_{s',s-1}~, \\
f_{\kappa\kappa'}^{+1}&=&{2\omega_{\kappa'\kappa}\over\omega_{Be}}
[(N+s+1)\delta_{N'N}+N\delta_{N',N-1}]\ |Z_{\kappa'\kappa}^{(0)}|^2
\delta_{s',s+1}~,   \\
f_{\kappa\kappa'}^0&=&{\hbar\omega_{\kappa'\kappa}\over{\rm Ry}}
\delta_{s's}\delta_{N'N} |Z_{\kappa'\kappa}^{(1)}|^2~,
\end{eqnarray}
where
\begin{equation}
Z_{\kappa'\kappa}^{(0)}=\int_{-\infty}^\infty {\rm d}z\
g^\ast_{\kappa'}(z)
g_\kappa (z)
\qquad {\rm and}\qquad
Z_{\kappa'\kappa}^{(1)}=\int_{-\infty}^\infty {\rm d}z\
g^\ast_{\kappa'}(z)
(z/a_B) g_\kappa (z)
\end{equation}
are the overlapping integral and dimensionless matrix element of the
longitudinal coordinate, respectively; $a_B$ is the Bohr radius.
The transition frequency is
\begin{equation}
\omega_{\kappa'\kappa}=(N'-N)\omega_{Be} + (N'-N+s'-s)\omega_{Bp}+
\tilde\omega_{\kappa'\kappa}~,
\end{equation}
where
 $\hbar\tilde\om_{\kappa'\kappa}=\epsilon_{\kappa'}-\epsilon_\kappa$
is the difference of the longitudinal energies,
and $\om_{Bp}=(m_e/m_p)\om_{Be}$ is the proton cyclotron frequency.
   In the application to the neutron star atmospheres, the only  important
case
is when
 the initial state $\kappa$ belongs
to the ground Landau level $N=0$ (the excited Landau levels are unpopulated
at $kT\ll\hbar\om_{Be}$). Moreover, we will consider  temperatures
low enough that the   population   of even the first excited level (010)
is small
compared to the ground state, (000),
which allows us to neglect transitions between the excited levels.
Then, putting $\kappa =(000)$, we obtain
\begin{eqnarray}
f_{\kappa\kappa'}^{-1}& =& 2{\omega_{Be}+\tilde\omega_{1-1n',000}
\over\omega_{Be}}|Z_{1-1n',000}^{(0)}|^2\delta_{N'1}\delta_{s'-1}~,
\label{fminus}
\\
f_{\kappa\kappa'}^{+1}& =& 2{\tilde\omega_{01n',000}+\omega_{Bp}\over
\omega_{Be}}
|Z_{01n',000}^{(0)}|^2\delta_{N'0}\delta_{s'1}~, \label{fplus}
\\
f_{\kappa\kappa'}^0 & = & {\hbar\tilde\omega_{00n',000}\over{\rm Ry}}
|Z_{00n',000}^{(1)}|^2 \delta_{N'0}
\delta_{s'0}~. \label{fzero}
\end{eqnarray}
Note that the longitudinal
energies and wave functions of the states $(01n)$ and (1--1$n$)
coincide (Hasegawa \&
Howard 1961). This means that $\tilde\omega_{1-1n',000}=
\tilde\omega_{01n',000}$
and $Z_{1-1n',000}^{(0)} = Z_{01n',000}^{(0)}$.
Since the parity of the longitudinal states is $(-1)^n$,  transitions
from the ground (even) state are allowed to odd states, $n=1,3,\ldots$, for
the longitudinal polarization $\al =0$, and to even states, $n=0,2,\ldots$,
for the circular polarizations $\al =\pm 1$.
Moreover, in the strong magnetic
field, the overlapping integrals $Z_{01n',000}^{(0)}$
for $n'\neq 0$ are much smaller than
that for $n'=0$ and can be neglected.
As a result, equations (\ref{fminus}) -- (\ref{fzero})
take the form
\begin{eqnarray}
f_{\kappa\kappa'}^{-1}& =& 2\left(1+{\tilde\omega_{010,000}\over\omega_{Be}}
\right) \delta_{N'1} \delta_{s'-1} \delta_{n'0}~, \label{fminus1} \\
f_{\kappa\kappa'}^{+1}& =& 2\left({\tilde\omega_{010,000}\over\omega_{Be}}
+{m_e\over m_p}\right) \delta_{N'0} \delta_{s'1} \delta_{n'0}~,
\label{fplus1}  \\
f_{\kappa\kappa'}^0& =& {\hbar\tilde\omega_{00n',000}\over{\rm Ry}}
|Z_{00n',000}^{(1)}|^2 \delta_{N'0} \delta_{s'0}~; \qquad n'=1,3,...
\label{fzero1}
\end{eqnarray}
These equations demonstrate that the  bound-bound transitions
are quite different for different polarizations. The strongest
transition (with $f_{\kappa\kappa'}^-\simeq 2$, $\omega_{\kappa'\kappa}
\simeq \omega_{Be}$) occurs onto the
excited Landau level, (000) $\to$ (1--10), and it is allowed only
for the left polarization. If the
electrons were not bound to protons by the Coulomb forces,
$\tilde\omega_{\kappa'\kappa}\to 0$,
this transition would turn into the usual electron cyclotron
absorption. Although the absorption in the left polarization
is negligible for the most interesting range $\omega\ll\omega_{Be}$,
we shall see below that it plays a crucial role for correct
evaluation of $\chi_{\pm 1}^H$ and of the NM polarizations in the
low-frequency range.
The only important transition for the
right polarization occurs onto the first excited tightly bound
level $(010)$; its oscillator strength, $f_{\kappa\kappa'}^+
\sim 2[\gamma^{-1} \ln\gamma  + (m_e/m_p)]$,
and  transition energy, $\hbar\omega_{\kappa'\kappa}\sim
{\rm Ry} \ln\gamma +\hbar\omega_{Bp}$, are  much smaller
than those for $\alpha=-1$. This transition would be  the
proton cyclotron absorption if  Coulomb binding could be
neglected.
The transitions allowed for the longitudinal polarization
proceed onto the hydrogen-like levels of odd parity,
creating a series of spectral lines with the leading
line formed by the transition to the lowest hydrogen-like
level, $(000)\to (001)$.
The oscillator strengths, $f_{\kappa\kappa'}\lapr 1$,
and the transition frequencies, $\hbar\omega_{\kappa'\kappa}
\sim {\rm Ry} \ln\gamma$, are smaller than
for $\alpha=-1$ but substantially greater than for
$\alpha=+1$. These transitions,  solely caused by
the Coulomb binding, disappear when the binding is
 turned off (if the electron-ion collisions are
neglected).

Equations~(\ref{fminus1})--(\ref{fzero1}) are obtained
in the adiabatic approximation, which leads to some errors
in the oscillator strengths.
In particular, this approximation
implies that $\tilde\om_{\kappa'\kappa}\ll \om_{Be}$ so that one cannot
trust second term in the numerator of (\ref{fminus1}).
Indeed, the slightly different
approach (still in the frames of the adiabatic
approximation) used by Hasegawa \& Howard (1961) leads
to the minus sign before
this term, instead of plus. Moreover, equations (\ref{fminus1}) and
(\ref{fplus1}) do not exactly satisfy the sum rule for the circular
polarizations (see, e.~g., Hasegawa \& Howard 1961),
\be
\sum_{\kappa'} \left(f_{\kappa\kappa'}^{-1}+f_{\kappa\kappa'}^{+1}
\right)=
2~.
\label{sumrul}
\ee
Indeed, the sum of the oscillator strengths
(\ref{fminus1}) and (\ref{fplus1}) for the  two
transitions, even without account for, e.~g., bound-free transitions
for the circular polarizations (see below), exceeds $2$.
 Therefore, more accurate values for the oscillator strengths,
obtained  numerically with allowance for the non-adiabatic
corrections,
should be used for precise calculations. We shall use below
the values from Forster et al.~(1984). For the magnetic
field $B=2.35\times 10^{12}$~G ($\gamma=1000$) and for the
transitions included into our calculations, they
are listed in Table~1.

 For the {\it bound-free} transitions, the final state
belongs to the continuum:
$\kappa'=(N's'\epsilon'\nu')$, where $\epsilon'$ ($>0$)
is the electron energy in
the (one-dimensional) continuum, and $\nu'=\pm$ is the parity of the
longitudinal function.  The sum over the final states in
equation~(\ref{abscoef})
is
\begin{equation}
\sum_{\kappa'} ={L_z\sqrt{2m_e}\over 2\pi\hbar}
\sum_{N's'\nu'}\int_0^\infty {d\epsilon'\over\sqrt{\epsilon'}}~,
\end{equation}
where $L_z$ is the the $z$-extension of the periodicity length
(the normalization length) of the continuum wave function
of the final state.
The photoionization cross section is the sum of the partial
cross sections
\be
\sigma_\kappa^\alpha(\omega)=\sum_{N's'\nu'}\sigma_{\kappa\to
N's'\nu'}^\alpha(\omega)
\label{sumpart}
\ee
which differ from zero for frequencies above the photoionization thresholds,
$\omega_{N's'}^{(t)}\equiv \epsilon_{000}/\hbar + [N'(1+m_e/m_p)+
s'm_e/m_p]\omega_{Be}$ if only the ground state is populated.
 The photoionization cross sections for the atoms at rest
have been considered
by Potekhin and Pavlov (1993) whose approach and codes we
shall use for computations and analysis.
In the dipole approximation, only one term contributes to the
sum (\ref{sumpart}) for a given polarization:
$N'=0$, $s'=0$, $\nu'=-\,$ for $\alpha=0$;
$N'=1$, $s'=-1$, $\nu'=+\,$ for $\alpha=-1$; and $N'=0$, $s'=1$,
$\nu'=+\,$ for
$\alpha=+1$. Then the cross sections
can be expressed in a form similar to that for the bound-bound transitions:
 \begin{eqnarray}
 \sigma_\kappa^{-1}(\omega)& =
& \pi a_B^2 \alpha_F
 {2\omega\over\omega_{Be}}\,
|\tilde{Z}_{1-1\epsilon'+,000}^{(0)}|^2\,
\sqrt{\frac{{\rm Ry}}{\hbar (\om -\omega_t^{-1})}}~,   \\
 \sigma_\kappa^{+1}(\omega)& =
& \pi a_B^2 \alpha_F
{2\omega\over \omega_{Be}}\,
|\tilde{Z}_{01\epsilon'+,000}^{(0)}|^2\,
\sqrt{\frac{{\rm Ry}}{\hbar(\om -\omega_t^{+1})}}~,   \\
\sigma_\kappa^0(\omega)& =
& \pi a_B^2 \alpha_F
{\hbar\omega\over{\rm Ry}}\,
|\tilde{Z}_{00\epsilon'-,000}^{(1)}|^2\,
 \sqrt{\frac{{\rm Ry}}{\hbar(\om -\omega_t^0)}}~,
 \end{eqnarray}
where
$\alpha_F$ is the fine structure constant,
\be
\tilde{Z}_{N's'\epsilon'\nu',000}^{(k)}=
(L_z/a_B) \int_{-\infty}^\infty  {\rm d}z\, g_{N's'\epsilon'\nu'}^\ast (z)
\, (z/a_B)^k\, g_{000}(z)\qquad (k=0,1)~,
\ee
$\hbar\omega_t^{-1}=\epsilon_{000}+\hbar\omega_{Be}$,
$\hbar\omega_t^{+1}=\epsilon_{000}+\hbar\omega_{Bp}$, and
$\hbar\omega_t^0 = \epsilon_{000}$ are the threshold energies,
and $\epsilon'=\hbar(\om - \om_t^\alpha)$ in the continuum wave
functions $g_{N's'\epsilon'\nu'}(z)$. Similar to the bound-bound transitions,
we have $\tilde{Z}_{1-1\epsilon'\nu',000}^{(0)}=
\tilde{Z}_{01\epsilon'\nu',000}^{(0)}$.
 It was shown by Hasegawa \&
Howard (1961) and Schmitt et al.~(1981) that the overlapping
integrals $\tilde{Z}^{(0)}$ and the matrix elements $\tilde{Z}^{(1)}$ are
proportional to $(\omega -\omega_t^\alpha)^{1/2}$ at
$\om\to\om_t^\alpha +0$,
so that the
cross sections are finite near the
thresholds. The overlapping integrals, and the
cross sections for the circular polarizations,
are very small because the
continuum wave functions are almost orthogonal to the wave
function of the ground state. A convenient measure for the
strength of the photoionization absorption for a given
polarization is the cumulative (integral) bound-free oscillator strength
obtained by integration over the final continuum states.
  From Table~1, where the integral bound-free oscillator
strengths are presented together with the photoionization thresholds,
we see that the bound-free transitions are much weaker than the
bound-bound ones for the circular polarizations, being somewhat
stronger than the bound-bound transitions for the longitudinal
polarization. Thus, the main transitions are:
$(000)\to (00\epsilon'-)$ and $(000)\to (001)$ for $\alpha=0$,
$(000)\to$(1--10)  for $\alpha=-1$, and $(000)\to (010)$ for $\alpha=+1$.
We include all the transitions listed in Table 1 into our
calculations of the absorption coefficients and
the components of the polarization tensor. Note that the
oscillator strengths in Table 1  satisfy the dipole sum rules with
accuracy certainly better than 1\%.

\subsection{Broadening Mechanisms and Effects of Finite Atomic Velocities }

The above equations correspond to the case of isolated atoms at rest.
They yield infinitely sharp features in both the antihermitian and hermitian
parts of the polarizability. Thermal motion of  atoms and interaction
with surrounding particles broaden and smooth the sharp features.
It was shown by Pavlov and M\'esz\'aros (1993) that the most important
broadening mechanism is associated with
different coupling of the internal and center-of-mass motions
(the difference between
transverse masses, in the perturbation
approach) of different levels.
In particular,
an infinitely narrow bound-bound profile
$\delta(\omega-\omega_{\kappa'\kappa})$ turns,
in the perturbation approach, into
\begin{equation}
\delta(\omega-\omega_{\kappa'\kappa})  \to
\psi_M(\omega-\omega_{\kappa'\kappa}) \equiv \left\{
               \begin{array}{ll}
                (\Gamma_M)^{-1}\exp\left[{\omega-\omega_{\kappa'\kappa})/
                                 \Gamma_M}\right]
                               & \omega<\omega_{\kappa'\kappa} \\

                  0 & \omega>\omega_{\kappa'\kappa}
               \end{array} \right. \label{magwid}
\end{equation}
The magnetic width $\Gamma_M$ is
\begin{equation}
\Gamma_M = {kT\over\hbar}\,
\left( 1 - {M^\perp_\kappa\over M^\perp_{\kappa'}}\right)~,
\end{equation}
where $M^\perp_\kappa$ is the transverse mass of the level $\kappa$.
The magnetic width grows linearly with temperature ($\hbar\Gamma_M
\simeq kT$ for the transitions from the ground level to highly
excited states) and dominates
all  other broadening mechanisms. The magnetic broadening
also smoothes the photoionization edges:
\begin{equation}
 \mu_\alpha^{\rm bf}(\omega) = n_a [1-\exp(-\hbar\omega/kT)]
\int_{{\rm max}(\omega,\omega_t^\alpha)}^\infty
{\rm d}\omega'\, \psi_M(\omega-\omega')\,
 \sigma_\kappa^\alpha(\omega')~,
 \label{bfabs}
\end{equation}
so that the bound-free profiles acquire exponentially decreasing
low-frequency tails.

Although the magnetic width
is very large, we see from equation~(\ref{magwid}) that it still retains
infinitely sharp
high-frequency edges
of the bound-bound profiles.
In order to avoid logarithmic singularities in $\chi^H$, arising after
the KK transformation of such functions, one should take into
account additional broadening mechanisms: Doppler broadening and/or
collisional broadening.

The Doppler width of the spectral lines becomes anisotropic in the
presence of strong magnetic field (Pavlov and \Mesz\ 1993).
For a transition $\kappa\to\kappa'$, the Doppler width
is
\begin{equation}
\Gamma_D =\alpha_F\, \omega_{\kappa'\kappa}
\sqrt{{m_e\over M}\, {kT\over {\rm Ry}}}
\sqrt{\cos^2\theta +
{MM_\kappa^\perp\over (M_{\kappa'}^\perp )^2} \sin^2\theta}~,
\end{equation}
where $M=m_p+m_e$ is the usual ``longitudinal'' mass of the atom,
and $\theta$ is the angle between the magnetic field and the wave
vector.

Since the density in the neutron star atmospheres is rather high,
collisional broadening
may be the most significant. The theory of collisional
broadening in very strong magnetic fields has not yet been
developed. A particular case of the Stark broadening by
plasma electrons, which is expected to be the most important
in strongly ionized gases, has been considered by Pavlov
\& Potekhin (1995) in the impact approximation. In the case
of weakly ionized or non-ionized gases the broadening
is mainly caused by interactions of neutral particles
(e.~g., van der Waals interactions). We know from
the theory   of spectral lines in the
non-magnetic case (e.~g., Sobelman, Vainshtein \& Yukov 1981)
that the most important broadening mechanism for strong transitions in a
single-component
gas is so-called self-broadening (or resonant broadening).
It results in a Lorentz profile with a width that
can be estimated as
\begin{equation}
\Gamma_s \sim {e^2\over m_e \omega_{\kappa'\kappa} }
f_{\kappa\kappa'}\, n_a = 4 {{\rm Ry}\over\hbar\omega_{\kappa'\kappa}}\,
f_{\kappa\kappa'}\, (n_a a_B^3)\, {{\rm Ry}\over\hbar}~,
\end{equation}
where $f_{\kappa\kappa'}$ is the oscillator strength of the transition,
and $n_a$ is the number density of atoms.
We adopt the same estimate for  strongly magnetized atoms.
The self-broadening width becomes negligibly small for very weak
transitions. However, it seems obvious that the collisional
width cannot be smaller than $\Gamma_g = \Gamma_{g\kappa} +
\Gamma_{g\kappa'}$, where
\be
\Gamma_{g\kappa} \sim 2n_a \langle\sigma v\rangle \simeq 2n_a
(\sigma_{\perp\kappa}
\langle v_{\perp\kappa}\rangle +
\sigma_{\parallel\kappa}\langle v_\parallel
\rangle ) = {4\over\alpha_F}\, (n_aa_B^3){\langle\sigma v\rangle
\over a_B^2 c}\, {{\rm Ry}\over\hbar}~,
\ee
and $\sigma_{\perp\kappa} \sim 2(r_\kappa+r_0)(l_\kappa+l_0)$ and
$\sigma_{\parallel\kappa} \sim \pi(r_\kappa+r_0)^2$ are  ``geometrical''
transverse and longitudinal cross sections for a collision of
the atom in the quantum state $\kappa$  with a background atom
in the ground state.
The cylindrical radius of the atom can be estimated as
$r_\kappa\sim a_B [(|s|+1)/\gamma]^{1/2}$; its longitudinal extension is
$l_\kappa\sim 1.2 a_B (|s|+1)^{1/4}(\ln\gamma)^{-1}$ for the tightly bound
levels, or $l_\kappa\sim 1.6 a_B \{{\rm Int}[(n+1)/2]\}^2$ for
the hydrogen-like levels (Potekhin 1994).
 The transverse and longitudinal mean relative velocities
of the colliding atoms are
$\langle v_{\perp\kappa}\rangle
=(\pi kT/2\mu_\kappa)^{1/2}$ and $\langle v_\parallel\rangle
= 2(kT/\pi M)^{1/2}$, where $\mu_\kappa = M_\perp^\kappa M_\perp^0
/(M_\perp^\kappa +M_\perp^0)$ is the reduced mass.
In addition, the lines are
broadened by   radiative damping (natural broadening).
The corresponding
widths are negligible for the transitions within
the Landau level $N=0$, but the radiative width considerably exceeds the
collisional widths for the main transition in the left
polarization, (000)$\to$(0--10). For this transition we have
\be
\Gamma_r={4e^2\omega_{Be}^2\over 3m_ec^3}={8\over 3}\,\alpha_F^3
\gamma^2 {{\rm Ry}\over\hbar}~.
\label{gammarad}
\ee
The collisional and radiative broadening of an
infinitely narrow line results in
the Lorentz profile
\be
\psi_L(\omega-\omega_{\kappa'\kappa})= {\Gamma_L\over 2\pi}\,
\left[(\omega-\omega_{\kappa'\kappa})^2 + {\Gamma_L^2\over 4}\right]^{-1};
\qquad\quad \Gamma_L=\Gamma_s+\Gamma_g+\Gamma_r~.
\ee
An example of the values of the magnetic, Lorentz and Doppler widths
 is presented in Table~1 for the transitions of interest. We see
that the Doppler and Lorentz widths are generally comparable.
Nevertheless, we shall neglect   Doppler broadening in our calculations.
The convolution of the Doppler and Lorentz profiles
coincides with the Lorentz profile far from the line center
(at $|\omega -\omega_{\kappa'\kappa}| \gapr \Gamma_D$),
and the resolution in the calculations presented here
is not sufficient to describe the details near the line
cores.  Moreover, the above estimates of the
widths are very approximate,
and the primary object of introducing the
broadening is to avoid sharp edges in the
absorption coefficients. Thus, we assume that the bound-bound lines
are broadened by   magnetic and Lorentz broadening:
\begin{equation}
\mu_\alpha^{\rm bb}(\omega)=
\frac{2\pi^2 e^2 n_a}{m_e c}\left[1-\exp\left(-\frac{\hbar\omega}{kT}
\right)\right]
\sum_{\kappa'}f_{\kappa\kappa'}^\alpha
\int_\omega^\infty {\rm d}\omega'\, \psi_M(\omega -
\omega')\, \psi_L(\omega'-\omega_{\kappa'\kappa})~.
\label{bbabs}
\end{equation}

\section{RESULTS}

 \subsection{Components of the Polarizability Tensor}

As a representative example,
we  investigated the NM properties for conditions
typical for photospheres of
moderately old, isolated neutron stars (Pavlov et al.~1995ab):
$B=2.35\times 10^{12}$ G ($\gamma=1000$, $\hbar\omega_{Be}
=2000$~Ry,
$\hbar\omega_{Bp}=1.09$~Ry), $kT = 1$~Ry ($T=1.58\times 10^5$~K),
$\rho = 1$~g cm$^{-3}$ ($n_a a_B^3 =0.0885$).
The ionization energy of hydrogen in this field is
$15.32$~Ry. All the relevant bound-bound and bound-free
transitions, the oscillator strengths, and the widths
are listed in Table~\ref{transitions}.
Although the problem  of ionization equilibrium  in strongly
magnetized, dense hydrogen has not been satisfactorily solved so far,
preliminary estimates show that the ionization degree is expected
to be low at this relatively low temperature
and high density, which allows us to neglect  the ionized
component and to demonstrate qualitative features
of the NMs in a nonionized, strongly magnetized gas.

Three absorption coefficients
$ \mu_\alpha(\omega)$,
calculated as described in Sections 2 and 3, are shown in Figure~1, and the
corresponding antihermitian  components of \pt\
$\chi_\alpha^A(\omega) =(c/4\pi\omega)\mu_\alpha(\omega)$
 are shown
in Figure 2.  We see that the behavior of $\mu_\alpha$ and
$\chi_\alpha^A(\omega)$
is quite different for different $\alpha$. For
$\alpha=-1$, it has one very strong resonance
associated with the bound-bound transition (000)$\to$(1--10) and
centered very close to the electron cyclotron frequency. The
strongest absorption in the most interesting frequency range
$\om\ll\om_{Be}$ is that for
radiation polarized along the field ($\alpha=0$). It
appears as a broad resonance around $\approx 15$~Ry
with a fine structure near the maximum associated with the peaks
of the overlapped bound-bound and bound-free absorption. The
high-energy tail of the resonance is due
the photoionization $(000)\to (00\epsilon'-)$, and the
low-energy tail is caused by the magnetic broadening of the bound-bound
transitions [mainly, of $(000)\to (001)$].
Two features are seen in $\mu_{+1}(\omega)$ and $\chi_{+1}^A(\omega)$:
the broadened, asymmetric  spectral
line $(000)\to (010)$ and the (much weaker) photoionization
spectrum $(000)\to (01\epsilon'+)$.

If the gas were fully
ionized, the $\chi_{-1}^A$ component would be almost the same ---
the peak would shift from 2004~Ry to the electron cyclotron
resonance at 2000~Ry. The $\chi_{+1}^A$ component would turn
into a narrow peak at the proton cyclotron frequency, the width
of the resonance being determined by the proton-proton collisions
(free charged particles are not subject to   magnetic
broadening --- their ``transverse masses'' are infinite). The
longitudinal component $\chi_0^A$ would decrease monotonically
with $\omega$ (the only ``resonance'' lies at $\omega=0$
for the unbound particles),
describing the inverse bremsstrahlung due to
the electron-ion collisions. Thus, the Coulomb binding
essentially alters the antihermitian part of the
polarization tensor, especially for the longitudinal component.

We made the KK transformation of $\omega\chi_\alpha^A$
(see eq.~[\ref{kramkron}]) and obtained the hermitian components
$\chi_\alpha^H(\omega)$ which are
shown in Figure~2 for $\omega>0$.
The accuracy of the numerical KK transformation
was  tested with   a simple model of   fully ionized plasma.
  As expected,
the hermitian part has spectral features
associated with those of the antihermitian part.
Indeed, as follows from equations~(\ref{kram-general}) and~(\ref{kramkron}),
an absorption
 feature
of $\chi_\alpha^A(\omega)$ centered at a positive
 frequency $\omega_\alpha$,
generates the contribution to $\chi_\alpha^H(\omega)$ equal to
$\omega_{pa}^2f^\alpha [8\pi\omega_\alpha (\omega_\alpha-\omega)]^{-1}$
 if the distance $|\omega-\omega_\alpha|$ from the feature
is much larger than its effective width,
and $\exp(-\hbar\omega_\alpha/kT)\ll 1$.
Each spectral feature at the positive frequency $\omega_\alpha$ has its
``counterpart'' at the negative frequency $-\omega_{-\alpha}$;
e.~g., for the spectral line (000)$\to$(010) in the
right polarization ($\omega_{+1}=\omega_{010,000}$, $f^{+1}=
f_{000,010}^{+1}$) the
counterpart line is centered at $-\omega_{-1}=-\omega_{1-10,000}$, and
its oscillator strength is $f^{-1}=f_{000,1-10}^{-1}$.
This counterpart generates
$\omega_{pa}^2f^{-\alpha}[8\pi\omega_{-\alpha}
(\omega_{-\alpha}+\omega)]^{-1}$. Summing these two terms
with allowance for the property $\omega_\alpha f^{-\alpha}=
\omega_{-\alpha} f^\alpha$ (see eqs.~[\ref{complet}] and [\ref{fminf}]), we
obtain
the total contribution of the feature to the hermitian part:
\be
\chi_\alpha^{H{\rm (bb)}} (\omega)=
\frac{\omega_{pa}^2}{8\pi}\, \frac{f^\alpha +f^{-\alpha}}
{(\omega_\alpha-\omega)(\omega_{-\alpha}+\omega)}~.
\label{farfromres}
\ee
This expression generalizes the well-known result (e.~g.,
Davydov 1976) of the isotropic case, when $\omega_{-1}=
\omega_{+1}=\omega_0$ and $f^{-1}=f^{+1}=f^0$ are the frequency and
the oscillator strength of a given transition.
Notice that equation (\ref{farfromres}) describes also
the polarizability of a ``cold'', magnetoactive plasma (e.~g.,
Ginzburg 1970) if $\omega_{pa}=\omega_{pe}=(4\pi n_e e^2/m_e)^{1/2}$
is the electron plasma frequency, $\omega_{-1}=\omega_{Be}$,
$\omega_{+1}=\omega_{Bp}$, $\omega_0=0$, $f^{-1}=2$,
$f^{+1}=2m_e/m_p$, and $f^0=1$. In our case of the cold,
nonionized gas, $\chi_{\pm 1}$ is determined by the bound-bound
transitions, so that equation (\ref{farfromres})
immediately gives the
circular components of the hermitian part:
\be
\chi_{\pm1}^H(\omega)\simeq\frac{\omega_{pa}^2}{4\pi (\omega_-\pm
\omega)(\omega_+\mp \omega)}~,
\label{farfromres1}
\ee
where $\omega_+\equiv \omega_{010,000}$ and $\omega_-
\equiv\omega_{1-10,000}$.
The only difference from the corresponding components for
the fully ionized plasma is that we replace   $\omega_{Bp}$ by
$\omega_+$,  and $\omega_{Be}$ by $\omega_-$.

Interaction of the longitudinally polarized radiation with the cold
gas is mainly  determined by the spectral line (000)$\to$(001)
and the photoionization (000)$\to$ (00$\epsilon-$). The former
yields
\be
\chi_0^{H{\rm (bb)}}=\frac{\omega_{pa}^2 f_0}
{4\pi(\omega_0^2-\omega^2)}~,
\label{chiHbb-large}
\ee
where $f_0\equiv f_{000,001}^0$ and $\omega_0\equiv\omega_{001,000}$.
Accurate values for the bound-free contribution can be obtained
only numerically. A qualitatively correct behavior of $\chi_0^{\rm (bf)}$
can be obtained from an approximate model cross section:
\be
\sigma^0(\omega)\approx\sigma_t\,\, (\omega_t/\omega)^3\,\,
\Theta(\omega-\omega_t)~,
\ee
where $\omega_t$ is the threshold frequency and $\sigma_t$ is
the threshold  cross section.
Substituting this model expression into
equations (\ref{kramkron}) and (\ref{abscoef}),
we obtain
\be
\chi_0^{H({\rm bf})}\simeq\frac{\omega_{pa}^2 f_{\rm (bf)}^0}
{4\pi\omega^2} \left[\left(\frac{\omega_t}{\omega}\right)^2
\ln\frac{\omega_t^2}{|\omega_t^2-\omega^2|} - 1\right]~,
\label{chiHbf}
\ee
where $f_{\rm (bf)}^0$ is the integral oscillator strength
of the bound-free absorption. At $\omega\ll\omega_t$, the
polarizability resembles that for the bound-bound transition
with $\omega_0\to (3/2)^{1/2}\omega_t$, $f^0 \to (3/4)f_{\rm (bf)}^0$\,:
\be
\chi_0^{H({\rm bf})}\simeq \frac{3\omega_{pa}^2 f_{\rm (bf)}^0}
{16\pi (\frac{3}{2}\omega_t^2 -\omega^2)}~.
\label{chiHbf-small}
\ee
When $\omega$ approaches $\omega_t$, the polarizability
diverges logarithmically:
\be
\chi_0^{H{\rm (bf)}}\simeq \frac{\omega_{pa}^2 f_{\rm (bf)}^0}
{4\pi\omega_t^2}\ln\frac{\omega_t}{2|\omega_t-\omega|}~.
\label{chiHbf-reson}
\ee
At $\omega >\omega_t$, it first decreases, changes its sign at
$\omega\simeq 1.16\omega_t$, reaches a minimum at $\omega
\simeq 1.58\omega_t$, and approaches zero at $\omega\gg\omega_t$:
\be
\chi_0^{H{\rm (bf)}}\simeq\frac{\omega_{pa}^2 f_{\rm (bf)}^0}
{4\pi[\omega_t^2\ln(\omega^2/\omega_t^2) -\omega^2]}~.
\label{chiHbf-large}
\ee

Equations (\ref{chiHbf})
through (\ref{chiHbf-large})
 are not  applicable too close to $\omega_+$
(=5.14 Ry), $\omega_0$ (=14.4 Ry),
$\omega_t$ (=15.3 Ry), and $\omega_-$ (=2004 Ry), where the bound-bound or
bound-free polarizabilities diverge if broadening mechanisms
are not included. The magnetically broadened spectral line
generates   the following longitudinal component of
the hermitian part of the
polarizability (at $\exp(-\hbar\omega_0/kT)\ll 1$)
\be
\chi_0^{H{\rm (bb)}}=
\frac{\omega_{pa}^2 f_0}{8\pi\omega\Gamma_M}
\left[\exp\left(-\frac{\omega_0-\omega}{\Gamma_M}\right)\,
{\rm Ei}\left(\frac{\omega_0-\omega}{\Gamma_M}\right) -
\exp\left(-\frac{\omega_0+\omega}{\Gamma_M}\right)
{\rm Ei}\left(\frac{\omega_0+\omega}{\Gamma_M}\right)\right]~,
\ee
where Ei($x$) is the exponential-integral function.
At $|\omega_0-\omega|\gg \Gamma_M$ this equation turns into
equation~(\ref{chiHbb-large}) with a shifted central frequency:
\be
\chi_0^{H{\rm (bb)}}=\frac{\omega_{pa}^2 f_0}{4\pi[(\omega_0-\Gamma_M)^2-
\omega^2]}~.
\label{chiHbbexp-large}
\ee
Near the central frequency of the unbroadened line, the polarizability
diverges logarithmically:
\be
\chi_0^{H{\rm (bb)}}=
-\frac{\omega_{pa}^2 f_0}{8\pi\omega_0\Gamma_M}\left[\ln\frac{\Gamma_M}
{|\omega_0-\omega|} -C\right]~,
\label{bbcenter}
\ee
where $C\simeq 0.577$ is the Euler constant. Thus,
with frequency growing from zero, the
polarizability generated by a magnetically broadened,
isolated spectral line first grows, reaches its maximum at
$\omega\simeq \omega_0-1.35\Gamma_M$, changes its sign at
$\omega\simeq\omega_0-0.37\Gamma_M$, reaches its logarithmically
infinite minimum at $\omega=\omega_0$, and then grows steadily to
zero.

The magnetically
broadened circular components are given by similar equations,
\be
\chi_{\pm 1}^{H{\rm (bb)}}=
\frac{\omega_{pa}^2}{8\pi\omega}
\left[\frac{f_{\pm}}{\Gamma_{M\pm}}
\exp\left(\frac{\omega-\omega_{\pm}}{\Gamma_{M\pm}}\right)
{\rm Ei}\left(\frac{\omega_{\pm}-\omega}{\Gamma_{M\pm}}\right) -
\frac{f_{\mp}}{\Gamma_{M\mp}}
\exp\left(-\frac{\omega_{\mp}+\omega}{\Gamma_{M\mp}}\right)
{\rm Ei}\left(\frac{\omega_{\mp}+\omega}{\Gamma_{M\mp}}\right)\right]~,
\ee
which turn into equations~(\ref{farfromres1}) far from the frequencies
$\omega_\pm$.
In the most interesting region $\omega\ll\omega_{Be}$,
the left-circular component varies very smoothly,
\be
\chi_{-1}^{H{\rm (bb)}} =
\frac{\omega_{pa}^2}{4\pi\omega_- (\omega_+ +\omega)}~,
\ee
whereas the right-circular component shows a resonant behavior near $\omega_+$,
similar to the above-described behavior of the longitudinal component
around $\omega_0$.

Magnetic broadening of the bound-free polarizability does not
change the polarizability at $|\omega-\omega_t|\gg \Gamma_M$
(see eqs.~[\ref{magwid}],
[\ref{chiHbf-small}], and [\ref{chiHbf-large}]),
and it removes   the logarithmic divergence at~$\omega_t$:
\be
\chi_0^{H{\rm (bf)}} \sim
\frac{\omega_{pa}^2 f_{\rm bf}^0}{4\pi\omega_t^2}
\ln\frac{\omega_t}{\Gamma_M}\qquad {\rm at}\qquad \omega\simeq\omega_t~.
\ee

The logarithmic divergence of the magnetically broadened
bound-bound polarizability (see eq.~[\ref{bbcenter}]) is
removed by the additional Lorentz (or Doppler) broadening.
The polarizability near the line center can be estimated as
\be
\chi_0^{H{\rm (bb)}}\sim -\frac{\omega_{pa}^2 f_0}
{8\pi\omega_0\Gamma_M}\ln\frac{\Gamma_M}{\Gamma_L}
\ee
for the longitudinal component. A similar estimate is true
for the right component, with replacing $\omega_0\to\omega_+$
and $f_0\to f_+$.

The above equations completely describe qualitative properties
of the cyclic components of the polarizability tensor. The results
of numerical calculation are shown in Figure 2.

\subsection{Polarization Properties of the Normal Modes}

The NM polarizations can be conveniently
described by the parameters $q$ and $p$ defined
in equations~(\ref{qpdef}) and (12).
In wide frequency ranges far from the bound-bound resonances
and the bound-free thresholds, the hermitian part of the polarizability
substantially exceeds the antihermitian part,
so that $|q|\gg |p|$, and the polarizations
are determined by the parameter
\be
q= \tilde{q}\frac{\sin^2\theta}{2\cos\theta}~,
\qquad {\rm where}\quad\tilde{q}\simeq
\frac {2\chi_0^H - \chi_+^H - \chi_-^H}{\chi_+^H-\chi_-^H}~.
\ee
In the case of fully ionized plasma, we have
\be
\tilde{q}=\frac{\om^2(\om_{Bp}^2+\om_{Be}^2-\om_{Bp}\om_{Be})-
\om_{Bp}^2\om_{Be}^2}
{\om^3(\om_{Be}^2-\om_{Bp}^2)}~.
\ee
At $\om\gg\om_{Bp}$
only the electron component is important, and $\tilde{q}=
\om_{Be}/\om$ is a smooth function of frequency. At $\om\gg\om_{Be}$
the parameter $\tilde{q}\ll 1$, that is, the NMs are circularly
polarized in a wide range of angles $\theta$. With decreasing
frequency, $\tilde{q}$ increases (the polarization ellipses become
narrower) and reaches its maximum, $\tilde{q}_{\rm max}\simeq 2\om_{Be}/
(3^{3/2}\om_{Bp})\simeq 707$ at $\om =\sqrt{3}\om_{Bp}$,
where the NM polarization is virtually linear; then it sharply decreases,
crosses zero in the very vicinity of proton cyclotron
resonance, at $\om\simeq\om_{Bp}(1+2\om_{Bp}/\om_{Be})\simeq 1.0011
\om_{Bp}$ and tends to $-\infty$ ($\tilde{q}\simeq -\om_{Be}\om_{Bp}^2
/\om^3$ at $\om\to 0$). The fact that $\tilde{q}$ is very small near
the proton cyclotron resonance means that one should expect
a peculiar behavior of the NMs there; in particular, strong
non-orthogonality of the NMs at some angles. To analyze the NM
properties around the $q=0$ point, one needs to calculate the
parameter $p$ whose behavior depends on the absorption mechanism
operating at these frequencies.

The relatively simple behavior of $q$ and $p$ for the fully ionized
plasma is due to the absence of resonances of the longitudinal
component of the polarizability $\chi_0$ at finite frequencies.
For the non-ionized gas, these resonances do exist due to
Coulomb binding, which strongly complicates the picture
(see Fig.~3). Nevertheless, even in this case the NM properties
can be investigated qualitatively with the aid of approximate
expressions for $\chi_\alpha$ derived above. At frequencies
well above the photoionization threshold $\om_t$ ($\gapr 20 - 30$ Ry
in our example), we obtain from equations~(\ref{farfromres1}),
(\ref{chiHbf-large}), and~(\ref{chiHbb-large}),
\be
\tilde{q}=\frac{\om_-}{\om}\left[1+
\frac{f_0\omega_0^2 + f_{\rm bf}\omega_t^2\ln(\om^2/\om_t^2)}
{\om^2}\right]~.
\label{qhighfreq}
\ee
We see that the correction for the longitudinal absorption/refraction
increases the value of $\tilde{q}$ as compared to the fully
ionized case, i.~e., the NM polarization is more linear than
in the fully ionized plasma. At frequencies much lower than
 $\om_0-\Gamma_M$ ($\lapr 12$ Ry in our example),
and not too close to $\om_+$,
we obtain from equations~(\ref{farfromres1}), (\ref{chiHbf-small}),
and (\ref{chiHbb-large}),
\be
\tilde{q}=\frac{\om_-(\om_+^2-\om^2)}{\om}
\left[\frac{f_0}{(\om_0-\Gamma_M)^2-\om^2}+\frac{3f_{\rm bf}}{2
(3\om_t^2-2\om^2)}\right]~.
 \label{qlowfreq}
\ee
At frequencies much below $\om_+$, we have $\tilde{q}\propto \om_-/\om$,
as  at high frequencies (eq.~\ref{qhighfreq}),
but this ratio is now multiplied by
a small factor $\sim f_0\om_+^2/\om_0^2$. Due the Coulomb  binding,
the behavior of $\tilde{q}$
at these low frequencies is qualitatively different from that
in  the fully ionized plasma. The function
$\tilde{q}(\om )$ changes its sign near $\om_+$ and keeps falling
with increasing $\om$
until $\om$ approaches $\omega_0-\Gamma_M$. Equation
(\ref{qlowfreq}) is not accurate near $\om_+$, where absorption plays
a role and the antihermitian component $\chi_{+1}^A$ should be
taken into account. Nevertheless, it satisfactorily describes the
qualitative behavior of $q$ even in this region.

It is clear from equations (\ref{qhighfreq}) and (\ref{qlowfreq})
that $q(\om )$ must change its sign at least once more, in a
region between $\sim (\om_0-\Gamma_M)$ and $\sim \om_t$. The behavior
of $q(\om )$ in this region is determined by interplay of the bound-bound
and bound-free refraction and absorption (both
hermitian and antihermitian components
are important), the details depending on the broadening mechanisms
and the value of the magnetic field.
Since $\tilde{q}\simeq 2\chi_0^H/(\chi_{+1}^H-\chi_{-1}^H)$
in this region,
and $\chi_{+1}^H-\chi_{-1}^H$ $(<0)$ is a smooth function of
frequency, the shape of $q(\om )$ reproduces that of $-\chi_0$
(cf.~Fig.~2). In particular, near the center of
an isolated line,
$\tilde{q}=\tilde{q}_{\rm min}\simeq -0.4 f_0\om_-/\Gamma_M$ at
$\om\simeq \om_0-1.4\Gamma_M$, and $\tilde{q}=\tilde{q}_{\rm max}
\sim (\om_-/2 \Gamma_M)\ln(\Gamma_M/\Gamma_L)$ at $\om\simeq\om_0$.
In our example, $\tilde{q}$ reaches its minimum, $\simeq -237$, at
$\hbar\om = 13.35$ Ry, then sharply grows, changes its sign at
$\hbar\om = 14.22$ Ry, reaches the maximum, $\simeq 472$, at
$\hbar\om = 14.42$ Ry,  and  then, after some ups and downs
associated with the longitudinal bound-free transition,
approaches the asymptotic curve (\ref{qhighfreq}).

The parameter $p$, which is associated with absorption of radiation
(antihermitian components), is important whenever $|p|$ exceeds
$|q|$, that is, in a very narrow range near  $\omega_+$,  and
in the range of strong longitudinal absorption ($\sim 13 - 20$
Ry in our example).
Around $\om_{+}$, the antihermitian part
is determined by the right-polarization absorption, $\chi_{+1}^A
\gg\chi_0^A, \chi_{-1}^A$, and $\chi_{+1}^H\gg\chi_{-1}^H$. In
this range
\be
\tilde{p}\equiv p\frac{2\cos\theta}{\sin^2\theta}\simeq -\frac{2\chi_{+1}^A
(\chi_0^H-\chi_{+1}^H)}{(\chi_{+1}^H)^2
+(\chi_{+1}^A)^2}~.
\ee
The ratio $\chi_{+1}^A/[(\chi_{+1}^H)^2+\chi_{+1}^A)^2]$ is a smooth
function of frequency (the resonances in the numerator and denominator
cancel each other), and so is  $\chi_0^H$. In our example,
$\chi_0^H$ exceeds $|\chi_{+1}^H|$ in this frequency range
(see Fig.~2),
and $\tilde{p}$ decreases with decreasing $\omega$ almost
smoothly, with a weak wiggle  around $5.1$~Ry caused by nonmonotonic
behavior of the relatively small $\chi_{+1}^H$. A characteristic
value of $\tilde{p}$ in this range can be roughly estimated
as $\tilde{p}\sim f_0\om_-\Gamma_{M+}/(\om_0^2-\om_+^2)$ ($\sim 1$
in our example --- see Fig.~3.).

In the range of strong longitudinal absorption
(and, more generally, at $\om_+\ll\om\ll\om_-$), we can neglect
$\chi_{\pm 1}^A$, so that
\be
\tilde{p}\simeq\frac{2\chi_0^A}{\chi_{+1}^H-\chi_{-1}^H}\simeq
-\frac{2c\, \om_-\, \mu_0(\om )}{\om_{pa}^2}~,
\ee
 and the dependence $p(\om )$ reproduces that of the
longitudinal absorption coefficient $\mu_0(\om )$. In
particular,  characteristic values of $\tilde{p}$ near
$\omega_0$ are $\sim f_0\om_-/\Gamma_M\sim\tilde{q}$ ($\sim 500$
in our example). They are considerably greater than those
near $\om_+$, which results in different behavior of the
NM polarizations near these frequencies.

In a wide range of  frequencies, where $|q|\gg |p|$
($\lapr 3$ Ry, between $\simeq 6$ and $\simeq 10$ Ry, and
at $\gapr 30$ Ry, in our example), the NMs are nearly
orthogonal:
\be
\delta_j=0, \pm\frac{\pi}{2};\qquad
{\cal{P}}_j=\pm\frac{1}{|q|+(q^2+1)^{1/2}}=\pm\frac{2\cos\theta}
{|\tilde{q}|\sin^2\theta+(\tilde{q}^2\sin^4\theta +4\cos^2\theta)^{1/2}}~.
\ee
In the most important domain,
$\om\ll\om_-$, we have $|\tilde{q}|\gg 1$ in the
wide frequency range, i.~e., the NMs are linearly polarized,
$|{\cal P}_j|\ll 1$,
at $\sin\theta \gg (2/\tilde{q})^{1/2}$. In the ranges of strong
longitudinal or right-polarization absorption, $q$ and $p$ become
comparable, and behavior of the NM polarizations is more
complicated. The most peculiar behavior occurs near the {\em critical
frequencies}
where $q=0$ ($\hbar\om_{c1}=4.751$~Ry and
$\hbar\om_{c2}=14.22$~Ry in our example).
For each of these frequencies, there exists an angle $\theta_c$
such that $|\tilde{p}(\omega_c)|\sin^2\theta_c/(2\cos\theta_c)
=1$;  the NM polarizations
at $\om =\om_c$ and $\theta=\theta_c$
are completely linear, ${\cal{P}}_j=0$, and
coincide with each other, $\delta_1=\delta_2=\pm \pi/4$
(complete non-orthogonality of NMs). In our example,
$\tilde{p}(\om_{c1})=-3.63$ and $\tilde{p}(\om_{c2})=-562$,
so that the non-orthogonality angles are
$\theta_{c1}\simeq 40.35^\circ$
and $\theta_{c2}=3.41^\circ$.
Since $|p(\om_c,\theta)|<1$ at $\theta<\theta_c$,  the circular
component of polarization is present, ${\cal{P}}_j=\pm
[(1-|p|)/(1+|p|)]^{1/2}$,
but the orientations of the polarization ellipses coincide,
$\delta_1=\delta_2=\pm \pi/4$. If $\theta > \theta_c$
(i.~e., $|p(\om_c,\theta)|>1$),
then the NMs are linearly polarized, ${\cal{P}}_j=0$, but
the polarization directions neither coincide nor are orthogonal,
$\tan 2\delta_j = \pm (p^2-1)^{-1/2}$.  The non-orthogonality points
and the curves $|q|=1$ and $|p|=1$ in the $\omega$-$\theta$ plane
 are shown in Figure 4.

In Figure  5  we  demonstrate the frequency dependence of the
 ellipticities and position angles
for three angles of propagation, $\theta=1^\circ,\, 30^\circ$, and
$75^\circ$, so that we cover three regions:
$\theta<\theta_{c2}$, $\theta_{c2}<\theta<\theta_{c1}$,
and $\theta>\theta_{c1}$.
 In the case of propagation nearly along the magnetic field,
the NMs are polarized circularly. The ellipticity ${\cal P}_j$ decreases
slightly in the range around $15$~Ry where $p$ is comparable   to $q$.
The position angles $\delta_j$ coincide at the two critical
frequencies $\om_{c1}$ and $\om_{c2}$.  Just above the second
critical frequency
the position angle varies strongly with frequency
due to the presence of bound-bound structure in $p$ (see Fig.~3).
For the intermediate angle
$\theta=30^\circ$,
the polarization of
the NMs becomes linear (${\cal P}_j=0$)
at the second critical frequency, while at the first one it is elliptical
with coinciding positional angles. The maximum ellipticity
 around $5$~Ry is due to the fact that in this region
$p$ is comparable to $q$, which increases the circular component of
polarization.
The polarizations of the NMs for  $\theta=75^\circ >\theta_{c1}$ are close to
linear.
They become exactly linear at the two critical frequencies
$\om_{c1}$ and $\om_{c2}$. Around $\om_{c1}$, however, the polarization
of the NMs is elliptical, since there is still a significant
range where $|q|$ is small and
$|p|$ is
close to unity.
This is also the cause of the small non-orthogonality of the NMs
around $\om_{c1}$.
 Around $\om_{c2}$, the NMs are polarized linearly and are orthogonal
because
$|p|$ is large and the curve $q(\omega)$
passes the range  $|q|\ll 1$ very rapidly.

The general picture is strikingly different from
the case of fully ionized plasma,
in which only one
critical frequency exists
in the vicinity of $\omega_{Bp}$. The Coulomb binding
not only shifts this point to higher frequencies and
changes the corresponding value of $\theta_c$,
but also creates at least one more
critical frequency in the range of strong longitudinal absorption.
Furthermore, it diminishes $|{\cal{P}}_j|$ at high frequencies,
$\gapr 30 $~Ry, and increases it at low frequencies,
$\lapr 5$~Ry.

The location of the critical points
where the polarization modes
acquire the above-described peculiarities
is a sensitive
function of the temperature and density because it depends strongly on the
shape of the absorption features.

\subsection{Absorption Coefficients of the Normal Modes}


It follows from the above analysis that
in  the vicinities of the  two points
($\omega_{c1}$, $\theta_{c1}$) and ($\omega_{c2}$, $\theta_{c2}$)
in the $\omega$-$\theta$ plane,
the NMs are strongly non-orthogonal (see Fig.~4). Far from these
non-orthogonality domains, the NM absorption coefficients
$\mu_j$ ($j=1,2$)
can be expressed as linear combinations of the `basic' coefficients
$\mu_\alpha$ (e.~g., Kaminker et al.~1982):
\be
\mu_j=\sum_{\alpha=-1}^{+1} a_j^\alpha \mu_\alpha~,
\ee
where
\be
a_j^0=\frac{\sin^2\theta}{2}\left[1+(-1)^j \frac{q}{\sqrt{1+q^2}}\right]~,
\ee
\be
a_j^{\pm 1}=\frac{1}{4}\left[1+\cos^2\theta \pm (-1)^j\frac{2\cos\theta
-q\sin^2\theta}
{\sqrt{1+q^2}}\right]~.
\ee
As we see from Figure 4, $|q|\gg 1$
(the NM polarizations are nearly linear) in a  wide domain of angles $\theta$
and frequencies $\omega\ll\omega_{Be}$.
For this domain
we obtain:
\be
\mu_1=\frac{\mu_{+1}+\mu_{-1}}{2} +
\frac{\mu_0\cos^2\theta}{\tilde{q}^2\sin^2\theta};
\qquad \mu_2=\mu_0\sin^2\theta + \frac{\mu_{+1}+\mu_{-1}}{2}\cos^2\theta~.
\label{quasitransv}
\ee
The  `ordinary mode' $j=2$ is polarized in the $\bf B$-$\bf q$ plane.
For purely transverse propagation, $\theta=90^\circ$, it is polarized
along the magnetic field, so that $\mu_2=\mu_0$. Since $\mu_{-1}\ll
\mu_{+1}$ at $\omega\ll\omega_{Be}$ (see Fig.~1), the admixture
of the transverse-polarization absorption at $\cos\theta\neq 0$ is
determined by the right-circular absorption coefficient $\mu_{+1}$.
The admixture term is negligibly small in the domain where equations
(\ref{quasitransv}) are applicable,   except for  a narrow frequency
range around the resonance of $\mu_{+1}$ ($\sim 2-5$ Ry in our case).
The `extraordinary mode' $j=1$ is polarized perpendicular to the
magnetic field, and $\mu_1=\mu_{+1}/2$ at $\theta=90^\circ$.
Since $\mu_0\gg\mu_{+1}$ in a wide frequency range,
the second term in the equation for $\mu_1$,
which describes the admixture of the longitudinal-polarization absorption,
may substantially exceed the first term at $\cos\theta\neq 0$, in spite of
the fact that it is proportional to a small factor $\tilde{q}^{-2}$.

At $\sin\theta \ll \tilde{q}^{-1/2}$ in a wide frequency range,
and at $\sin\theta\ll\sin\theta_c$ near the critical frequencies,
the NM polarizations
are nearly circular ($|q|\ll 1$ and $|p|\ll 1$). In this domain,
\be
\mu_1=\mu_{-1}\left[\frac{(1+\cos\theta)^2}{4}-\frac{\tilde{q}\sin^4\theta}
{8\cos\theta}\right]+\mu_0\frac{\sin^2\theta}{2}+
\mu_{+1}\left[\frac{(1-\cos\theta)^2}{4}+\frac{\tilde{q}\sin^4\theta}
{8\cos\theta}\right]~,
\label{quasilong1}
\ee
\be
\mu_2=\mu_{+1}\left[\frac{(1+\cos\theta)^2}{4}-\frac{\tilde{q}\sin^4\theta}
{8\cos\theta}\right] +\mu_0\frac{\sin^2\theta}{2} +
\mu_{-1}\left[\frac{(1-\cos\theta)^2}{4}+\frac{\tilde{q}\sin^4\theta}
{8\cos\theta}\right]~.
\label{quasilong2}
\ee
For purely longitudinal propagation, $\theta=0$, the NM polarizations
are exactly circular, so that $\mu_1=\mu_{-1}$ and $\mu_2=\mu_{+1}$.
The left-circular absorption is very weak; in fact, at $\om\ll\omega_{Be}$,
it is determined by violation of the dipole selection rules due to
the motion of atoms in the magnetic field (Pavlov \& Potekhin 1995), which
is neglected in our perturbation approach. However, even at very
small $\theta\neq 0$, absorption in the $j=1$ mode is mainly determined
by the admixture of the (relatively strong) right-circular and longitudinal
absorptions, $\mu_1\simeq (\mu_{+1}/8)\sin^4\theta + (\mu_0/2)\sin^2\theta$,
and the exact value of $\mu_{-1}$ is not important.

Equations (\ref{quasitransv}), (\ref{quasilong1}), and (\ref{quasilong2})
allow one to understand qualitatively behavior of the spectral
and angular dependences of $\mu_j$ depicted in Figures 6, 7, and 8.
%
%
Figure 6 demonstrates  evolution of  $\mu_j(\omega)$
with increasing $\theta$ for directions of propagation close to
the magnetic field.
For $\theta\ll\theta_{c2}$ ($\theta=1^\circ$ in Fig.~6),
 the NMs are polarized almost exactly circularly at all the frequencies.
The absorption coefficient of the $j=2$ mode (dashed curve) is determined
by the right-circular absorption $\mu_{+1}$
 (dominates at $\hbar\om
\lapr 8$ Ry [bound-bound transition to (010)] and $\gapr 15$ Ry
[bound-free transition to $(01\epsilon+)$] --- cf.~Fig.~1)
and by a small addition, $(\mu_0/2) \sin^2\theta$
of the longitudinal absorption (dominates at
the longitudinal resonance, $\hbar\om \sim 10-15$ Ry).
The other mode is solely determined (at this angle)
 by the same contribution of the longitudinal
absorption.
The absorption coefficients $\mu_1$ and $\mu_2$ almost coincide
with each other at the region of strong longitudinal absorption.

The increase of  $\theta$ towards and beyond the critical value
$\theta_{c2}=3.41^\circ$ is accompanied by fast growth of
both NM absorption coefficients in the region of prevailing
longitudinal absorption, $\hbar\om\gapr 10$ Ry; the peak at
$\hbar\om \simeq 14 - 15$ Ry becomes $\simeq 40$ times higher
when $\theta$ changes from $1^\circ$ to $5^\circ$. The coefficients
also approach each other at these frequencies when $\theta$ approaches
$\theta_{c2}$, and separate again when $\theta$ moves away from
$\theta_{c2}$ (see also Fig.~8). The resonance of right-circular
absorption around 5~Ry remains almost unchanged in one mode,
and its admixture to the other mode grows $\propto \sin^4\theta$.

Figure 7 demonstrates behavior of $\mu_j(\omega)$ at intermediate
$\theta$. The left panel ($\theta=30^\circ$) shows a typical
frequency dependence at
$\theta_{c2}< \theta < \theta_{c1}$.
The NM polarization is almost circular around the critical frequency
$\hbar\omega_{c1}=4.75$~Ry, and the absorption coefficients can be
evaluated there from equations (\ref{quasilong1}) and (\ref{quasilong2}).
The main distinction from the case of small $\theta$ is that the
right-circular absorption gives a larger contribution to $\mu_1$,
so that the resonance in that NM mode became much more pronounced.
At $\hbar\om\gapr 6$~Ry and $\lapr 4$~Ry, the NMs are polarized almost
linearly,
(see Figure 5), and their absorption coefficients
are given by equation (\ref{quasitransv}). It should be noted,
however, that $\mu_2$ in equation (\ref{quasitransv}) corresponds
to the solid line at $\hbar\om\gapr 6$~Ry, and to the dashed line
at $\hbar\om\lapr 4$~Ry. The reason is that in the presence of
the critical frequencies and non-orthogonality points a unique
  labeling of the NMs along the continuous curves of the NM
absorption and refraction is impossible (see Pavlov et al.~1980);
this fact, however, does not impose  any difficulties
because both NMs participate in the radiative transfer, and their
labeling is purely conventional.

The angle $\theta=41^\circ$ only slightly exceeds $\theta_{c1}=40.35^\circ$.
At this angle the NMs are substantially non-orthogonal near $\omega_{c1}$,
and the absorption coefficients nearly coincide in this region
of the right-circular resonance (the middle panel of Fig.~7). At frequencies
below and above this region, the NM polarization is close to linear,
and the NM behavior is similar to that for $\theta=30^\circ$.

At the angle $\theta=75^\circ$, which is well above $\theta_{c1}$, the NM
polarizations are almost linear at all frequencies, although there
is a very narrow region of enhanced ellipticity at $\hbar\omega\simeq 5.1$~Ry
(see Fig.~5).
The absorption coefficients are fairly accurately described by equation
(\ref{quasitransv}) ($\mu_2$ corresponds to the solid curve for all the
frequencies). Notice that $\mu_1$ and $\mu_2$ coincide just in the
enhanced ellipticity region.



Figure 8 presents the absorption coefficients $\mu_j$ as a function of
the propagation angle for a few characteristic frequencies.
At low frequencies $\om<\om_{c1}$ (see the panel for $\hbar\om=2$~Ry
in Fig.~8),
where the absorption coefficients are determined by the
magnetically broadened wings of the bound-bound lines,
 the NM polarization changes from circular to
linear with increasing angle. Absorption in one mode
slightly decreases with $\theta$, while in the other mode it
grows very fast at small angles because of increasing contribution of $\mu_0$
(see eq.~[\ref{quasilong2}]); the coefficients become close to each
other at $\theta\gapr 20^\circ$, where the NMs are almost linearly
polarized.

The picture changes considerably when we approach the first critical frequency,
$\hbar\om_{c1}=4.751$~Ry. Here, for $\theta=\theta_{c1}=40.35^\circ$, the
(linear) polarizations and the absorption coefficients of the modes
coincide with each other: $\delta_1=\delta_2=\pi/4$ and
${\cal P}_1={\cal P}_2=0$. The modes are linearly
polarized at angles above $\theta_{c1}$,
and they become more orthogonal with increasing $\theta$. Below
$\theta_{c1}$, their polarizations are elliptical, and the ellipticity
increases
with decreasing angle.

In the region between the
critical frequencies, $\om_{c1} < \om < \om_{c2}$,
($\hbar\om=7$~Ry in Fig.~8),
the longitudinal absorption $\mu_0$
(the low-frequency
wing of the longitudinal resonance) is much greater than
the right-circular absorption $\mu_{+1}$
(the high-frequency
wing of the spectral line and the low-frequency
 wing of the photoionization edge).
The NM polarization is nearly circular at small $\theta$
($\lapr 10^\circ$ for $\hbar\omega=7$ Ry),
and, in accordance with equations (\ref{quasilong1}) and (\ref{quasilong2}),
both $\mu_1$ and $\mu_2$ grow with $\theta$:
$\mu_1\simeq \mu_{+1}+(\mu_0/2)\sin^2\theta$, $\mu_2\simeq (\mu_0/2)
\sin^2\theta$. The polarization becomes almost linear at $\theta\gapr
20^\circ$,
where $\mu_2\simeq \mu_0 \sin^2\theta$ keeps growing, whereas
$\mu_1=(\mu_{+1}/2) + (\mu_0/\tilde{q}^2)\cot^2\theta$ decreases because
of diminishing contribution from the longitudinal absorption.

At the second critical frequency $\hbar\om_{c2}=14.22$~Ry,
the NMs coincide at the critical angle $\theta_{c2}=3.41^\circ$.
For $\theta< \theta_{c2}$, the NMs modes are nearly circular,
and the absorption coefficients increase with angle,
being very close to each other (cf.~Fig.~6) because the both are mainly
contributed from the longitudinal absorption ($\mu_0/\mu_{+1}\simeq
1.3\times 10^5$ at this frequency).
For $\theta>\theta_{c2}$, the linearly polarized NMs become more
orthogonal with increasing $\theta$: $\delta_2\to 0$ and $\delta_1\to \pi/2$;
absorption coefficients of the ordinary and extraordinary modes
grow up to $\mu_0$ and decrease down to $\mu_{+1}/2$, respectively.

Behavior of $\mu_j(\theta)$
above the second critical  frequency
is demonstrated in Figure 8 for  $\hbar\om=14.82$~Ry,
near the peak of longitudinal absorption,
and for $\hbar\om=30$~Ry, where both $\mu_0$ and $\mu_1$ are determined by the
bound-free transitions.
At these frequencies, the NM polarizations are virtually orthogonal
for any $\theta$.
Since $|\tilde{q}|\gg 1$, the polarizations are circular in a narrow
range of angles, $\sin\theta\lapr \tilde{q}^{-1/2}$, and they are
linear in a wide range of angles. The absorption coefficient of the
ordinary mode grows monotonically with $\theta$: $\mu_2\simeq \mu_{+1}+
(\mu_0/2)\sin^2\theta$ at small $\theta$, and $\mu_2\simeq \mu_0\sin^2\theta$
at larger angles. For the extraordinary
mode, $\mu_1\simeq (\mu_0/2) \sin^2\theta$ grows at small $\theta$, reaches
its maximum value, $\sim \mu_0/\tilde{q}$ at $\sin\theta \sim
\tilde{q}^{-1/2}$,
and then decreases gradually, $\mu_1\simeq (\mu_0/\tilde{q}^2)\cot^2\theta +
\mu_{+1}/2$.
%

\section{DISCUSSION}

In this paper we
suggested  a convenient method for calculating the
absorption coefficients of the normal modes in a magnetized medium.
Given the `basic' absorption
coefficients
$\mu_0$ (for the linear polarization along $\bf B$ at transverse
propagation) and $\mu_{\pm 1}$ (for the circular polarizations at
longitudinal propagation),
we  obtain the antihermitian part
of the \pt\ and
use the Kramers-Kronig
transformation to calculate its   hermitian part.  The knowledge
of the full \pt\ allows us to calculate the polarizations
and the absorption coefficients of the NMs
for any frequency and direction of propagation. We applied the method
to a nonionized, strongly magnetized hydrogen gas for conditions
expected in atmospheres of neutron stars.
Absorption and propagation of radiation in such a medium is determined
by the bound-bound and bound-free transitions.  The
corresponding  spectral lines
and photoionization edges are strongly broadened due to distortion of the
structure of atoms moving across the magnetic fields,
and the  shapes of the lines
 depend essentially on the direction of propagation
and polarization of radiation.
As a result, the angular and spectral dependences of the NM characteristics
are much more complicated than in a  fully ionized plasma. In particular,
the presence of the nonionized component leads to appearance of
additional critical frequencies  and angles,  where peculiarities
of the NM polarizations cause additional crossings of the curves
$\mu_j(\omega)$ or $\mu_j(\theta)$.

The resulting absorption coefficients can be used to
solve the radiative transfer equations for the NMs
in neutron star atmospheres and to  calculate
spectra, angular distribution and polarization of
emergent radiation. Since the temperature grows inward in
the atmosphere, the spectrum   and direction  of
the radiation emitted is mainly determined by the NM which
has lower opacity and, therefore, escapes from deeper and
hotter layers. We see from  Figure 6 that at small $\theta$
the NM of lower opacity shows a strong, asymmetric resonance
associated with the bound-bound and bound-free transitions
for the
longitudinal polarization. Consequently, we should expect
a broad, asymmetric absorption feature (at $\hbar\om \sim 14-15$~Ry in
our example)
 in the emergent spectrum observed at directions close to the
magnetic field direction (e.~g., when the neutron star is seen
pole-on). Figure 7 shows that at intermediate and large $\theta$
the strongest absorption in the NM of lower opacity is that
at the bound-bound resonance for the right-circular polarization
(at $\sim 5$~Ry), so that the corresponding
absorption feature should be most clearly seen at these directions.
 From the angular dependences of $\mu_j$ shown in Figure 8
we see that at most frequencies
the NM of lower opacity
sharply grows with $\theta$ at small $\theta$, reaches a maximum,
and then decreases gradually when $\theta \to 90^\circ$. This means
that the angular distribution $I(\theta)$ of
the intensity of radiation  is beamed
along the magnetic field (a `pencil component' of the angular
distribution); a typical width of the beam is
$\Delta\theta\lapr \tilde{q}^{-1/2}$
at $\tilde{q}\gg 1$ ($\Delta\theta\sim 3^\circ - 10^\circ$
at $\hbar\om\sim 6 - 30$~Ry in our example). The maximum of
$\mu_j(\theta)$ should correspond to a minimum of $I(\theta)$,
i.~e., a broad `fan component' in the plane perpendicular
to the magnetic field is emitted in addition to the sharp `pencil component'.
The pencil component broadens when the frequency approaches $\om_{c1}$; at
$\om <\om_{c1}$ the minimum of $I(\theta)$ (and the fan component)
disappears. The polarization of
the emergent radiation is expected to be strongly polarized
at frequencies and angles where the the NM absorption coefficients
are substantially different. The polarization is circular in the
narrow pencil component, and it is linear (perpendicular to
the magnetic field) in the fan component.

In the present paper we considered only the case of low temperatures,
when the atoms are expected to be nonionized
and the effect of their motion on the radiative transitions
can be considered in the framework of the perturbation theory. At higher
temperatures, the ionized component of the plasma would contribute
to the polarization tensor, NM polarizations, and NM opacities.
This will result in even more complicated spectral and angular
behavior of the NM characteristics. To include both ionized and nonionized
components, one should first solve the nontrivial problem of
ionization equilibrium of
a strongly magnetized, highly nonideal plasma.
Including the (nonperturbative) effects of the atomic motion at higher
temperatures (Bezchastnov \& Potekhin 1994; Pavlov \& Potekhin 1995)
would be also desirable, particularly for propagation at directions
close to the magnetic field.
Besides,
it is
possible that not only hydrogen, but also other atoms
(e.g. helium, or even iron) can be present in
neutron star atmospheres. The study of the structure and radiative transitions
in the heavier atoms in very strong magnetic fields is a very complicated
problem. When these problems are solved, the results can be easily
used to calculate the NM properties with the
method suggested
in  the present paper.

The research has been partially supported through NASA grants
NAG5-2807, NAG5-2868 and NASW-4690.
 We are grateful to Alexander Potekhin who supplied
the codes for calculating the photoionization cross section
used in this paper. We thank Victor Bezchastnov, Cole Miller,
and Yura Shibanov for useful discussions.

%
%
%
\newpage
\begin{table}[ph]
\begin{center}
\begin{tabular}{cccccccc}
\tableline
\tableline
 Polari- & Final & Final  & Transition & Oscillator & & & \\
zation  & level & energy &  energy    & strength   &$\hbar\Gamma_M$ &
$\hbar\Gamma_L$ & $\hbar\Gamma_D$ \\
   $\alpha$      & $N's'n'$   &  [Ry]       &   [Ry] &  $f$
& [Ry]          &  [Ry]            &   [Ry]       \\
\tableline
  0         &  001         & $-0.99$     &  $14.40$   &   0.2272
& 0.987         &   0.0076
&  0.0024             \\
  0         &  003         & $-0.25$     &  $15.08$   &   0.0245
& 0.998         &   0.0065
& 0.0026              \\
  0         &  005         & $-0.11$    &  $15.21$     &  0.0071
& 0.999         &    0.013
& 0.0026             \\
  0        & $00\epsilon$  & $>0$        &  $>15.32$   &   0.74
& 1.00          &                  &              \\
 +1         &  010         & $-10.19$   &  $5.14$     &   0.0051
& 0.464        &    0.0014
& 0.0012             \\
 +1         & $01\epsilon'$ & $>1.09$    &  $>16.41$   &   0.0004
& 1.00          &                  &               \\
$-1$         & 1$-10$       & $1988.72$  &  $2004.05$  & $1.971$
&  1.00             &    1.037         &  0.341             \\
 $-1$        & 1$-1\epsilon'$ & $>2000.00$     & $>2015.32$   & $0.02$
& 1.00          &                  &                  \\
\tableline
\end{tabular}
\end{center}
  \caption[]{Bound-bound and bound-free transitions from the ground  state,
their oscilator strengths,
 and magnetic, Lorentz, and Doppler widths of the associated lines
and pho\-to\-ioniza\-tion edges at
$kT=1$~Ry, $\rho=1$~g~cm$^{-3}$, and $B=2.35\times 10^{12}$~G.}
\label{transitions}
\end{table}

%
%
%
\newpage

\section*{References}
{\parindent=0pt

Bezchastnov, V.~G., \& Potekhin, A.~Y. 1994, J.~Phys.~B,
27, 3349.

Canuto, V. \& Ventura, J. 1977, Fundam.~Cosmic Phys., 2, 203.

Davydov, A. S. 1976, {\em Quantum Mechanics}, Pergamon Press.

Forster, H., Strupat, W., R\"osner, W., Wunner, G., Ruder, H.,
\& Herold, H. 1984, J.~Phys.~B, 17, 1301.

Ginzburg, V. L. 1970, {\em The Propagation of Electromagnetic Waves in
Plasmas}, Pergamon Press.

Gnedin, Yu.~N., \& Pavlov, G.~G. 1974,  Sov.~Phys.~JETP, 39, 301.

Hasegawa H. \& Howard R. 1961, J.~Phys.~Chem.~Solids, 21, 179.

Herold, H., Ruder, H., \& Wunner, G. 1981, J.~Phys.~B, 14, 751.

Kaminker, A.~D,, Pavlov, G.~G., \& Shibanov, Yu.~A. 1982, Ap\&SS, 86, 249.

Kaminker, A.~D., Pavlov, G.~G., \& Shibanov, Yu.~A. 1983,
Ap\&SS, 91, 167.

Landau, L.~D., \& Lifshitz, E.~M. 1960, {\em Electrodynamics of Continuous
Media},
Pergamon Press.

\Mesz, P. 1992, {\em High Energy Radiation from Compact Objects}, University
of Chicago Press.

\"Ogelman, H. 1995, in {\em The Lives of the Neutron Stars}, eds.
A.~Alpar, \"U. Kizilo\v{g}lu and J.~van~Paradijs
(Dordrecht: Kluwer).

Pavlov, G.~G., Shibanov, Yu.~A., \& Yakovlev, D.~G. 1980, Ap\&SS, 73, 33.

Pavlov, G.~G., \& \Mesz, P. 1993,  ApJ, 416, 752.

Pavlov, G.~G., Shibanov, Yu.~A., Zavlin, V.~E., \& Meyer, R.~D. 1995,
in {\em The Lives of the Neutron Stars}, eds. A.~Alpar, \"U. Kizilo\v{g}lu
and J.~van Paradijs (Dordrecht: Kluwer)

Pavlov, G.~G., \& Potekhin, A.~Y. 1995, ApJ, 450, 883.

Pavlov, G.~G., Stringfellow, G.~S., \& C\'ordova, A.~F. 1995, ApJ, submitted.

Potekhin, A.~Y. 1994, J.~Phys.~B, 1994, 27, 1073.

Potekhin, A. and Pavlov, G.: 1993, Ap.J., 407, 330.

Schmitt, W., Herold, H., Ruder, H., \& Wunner, G, 1981, A\&A, 94, 194.

Sobelman, I,~I., Vainshtein, L.~A., \& Yukov, E.~A. 1981,
{\em Excitation of Atoms and Broadening of Spectral Lines}, Springer-Verlag.

Vincke, M., \& Baye, D. 1988, J.~Phys.~B, 21 2407.

Vincke, M., Le Dourneuf, M., \& Baye, D. 1992, J.~Phys.~B, 25, 2787.

} 

\newpage
\section*{Figure Captions}

Figure 1. Frequency dependences of the absorption coefficients $\mu_\alpha$
for the three
polarizations: $\alpha=0$ (linear polarization along the magnetic field $\bf
B$),
and $\alpha=\pm 1$ (circular polarizations in the plane perpendicular to $\bf
B$).
The right panel shows detailed structure of $\mu_0$ and $\mu_{+1}$
in the range of  bound-bound resonances and bound-free edges.
The contributions
of the bound-bound (bb) and bound-free (bf) transitions
are shown
separately. The low-frequency wings of the
resonances are  determined by the
magnetically broadened spectral lines.

Figure 2. The upper panel shows the circular components $\chi_{\pm 1}$ of the
hermitian and antihermitian parts
polarizability tensor. The left-circular component $\chi_-^A$ is very
small
in this frequency range.
The major features of $\chi_+^A$ and $\chi_+^H$  are associated
with the bound-bound resonance  at $5.1$~Ry.
 The lower panel shows $\chi_0^A$ and $\chi_0^H$. The main
features (shown in the inset separately) are due to the photoionization
edge at 15.3~Ry  and bound-bound resonance at 14.4~Ry.

Figure 3. The upper panel shows the global behavior of the
parameters $\tilde q$ and
$\tilde p$ which determine the NM polarization.
The two lower panels
show $\tilde q$ and $\tilde p$ in the vicinity of
the critical frequencies where
$\tilde q =0$. Note the different vertical scales.

Figure 4. The solid and dashed curves connect the points in the
$\omega$-$\theta$
plane for which $|q|=1$ and $|p|=1$, respectively.
The labels $c_1$ and $c_2$
denote the non-orthogonality points where $q=0$ and $|p|=1$.

Figure 5. The left three panels show ellipticity and the right
three panels show position angles of the NM polarization ellipses
as
functions of frequency for three angles between $\bf B$ and
direction of propagation:
$\theta=1^\circ,\; 30^\circ$, and $75^\circ$. Note the different scale in the
lower panels.

Figure 6. Absorption coefficients
of the NMs vs.~frequency
for
three directions of propagation close to $\bf B$.

Figure 7. Absorption coefficients of the NMs
vs.~frequency for three
intermediate angles $\theta$.

Figure 8. Angular dependences of the absorption coefficients of the NMs
for six
photon energies.
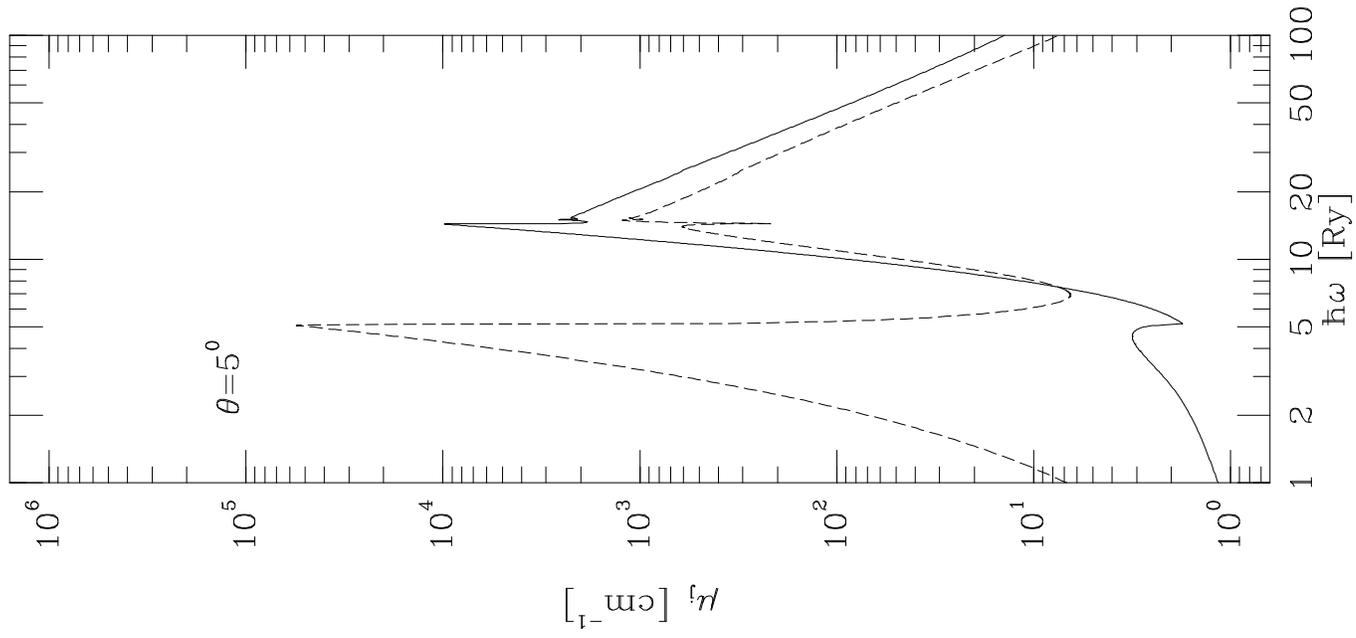
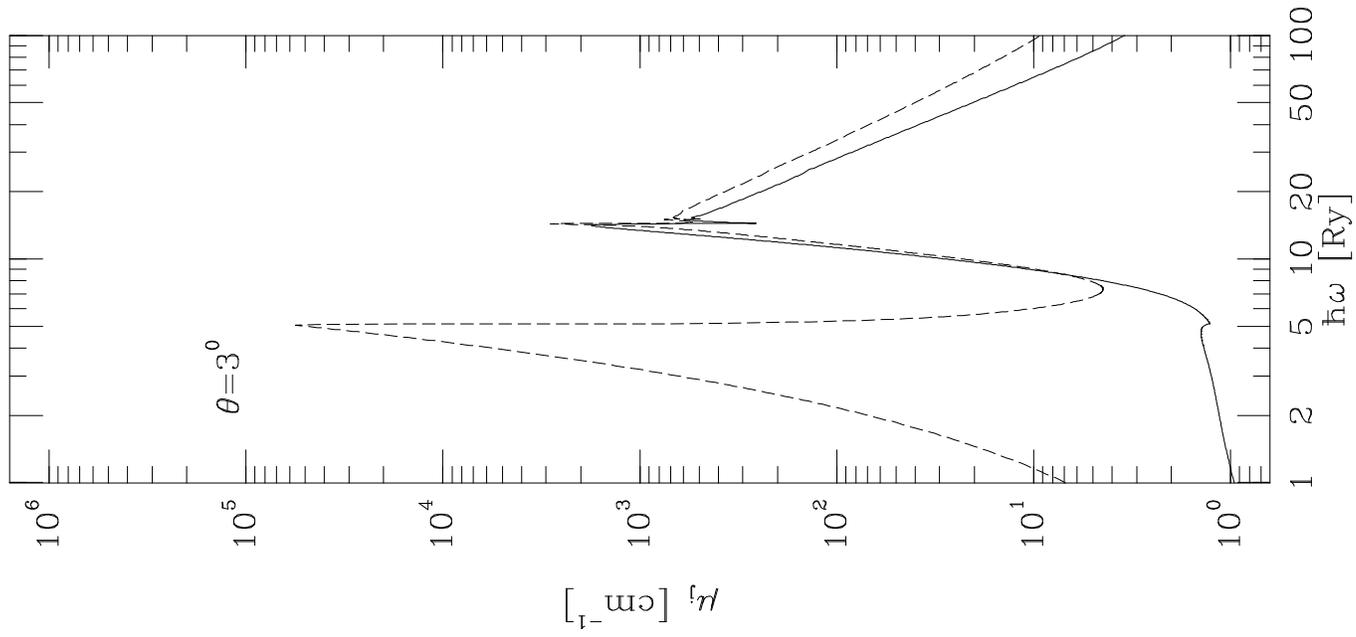
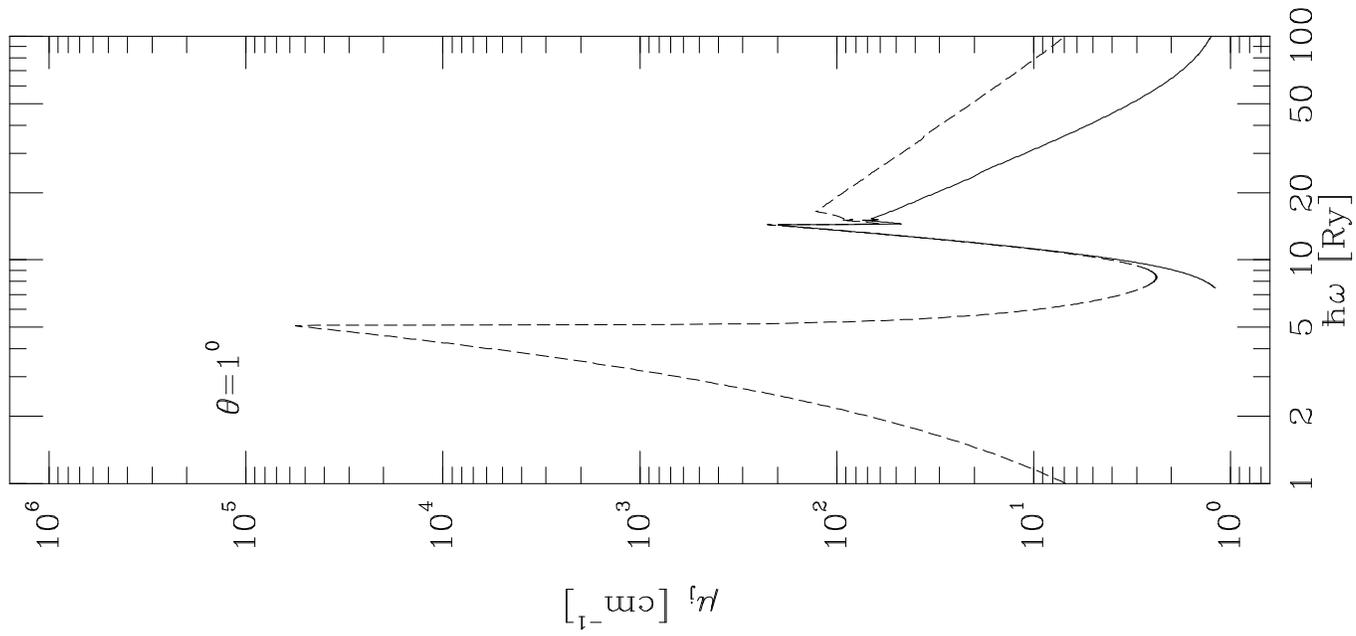
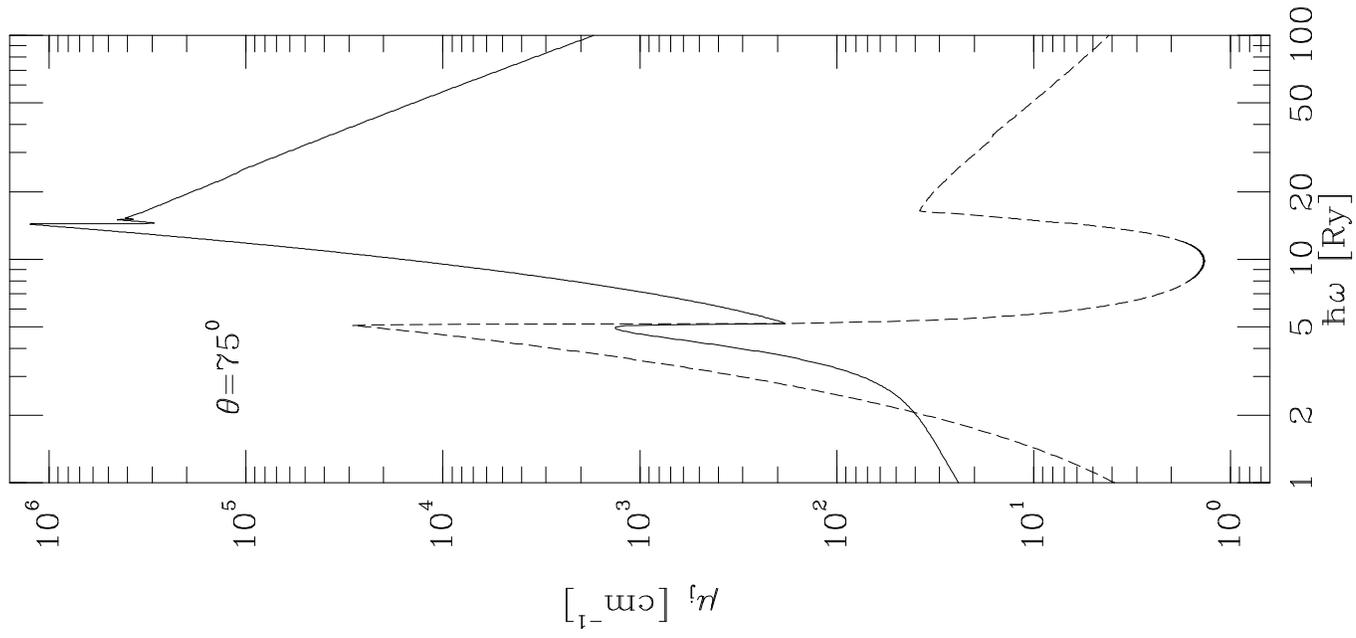
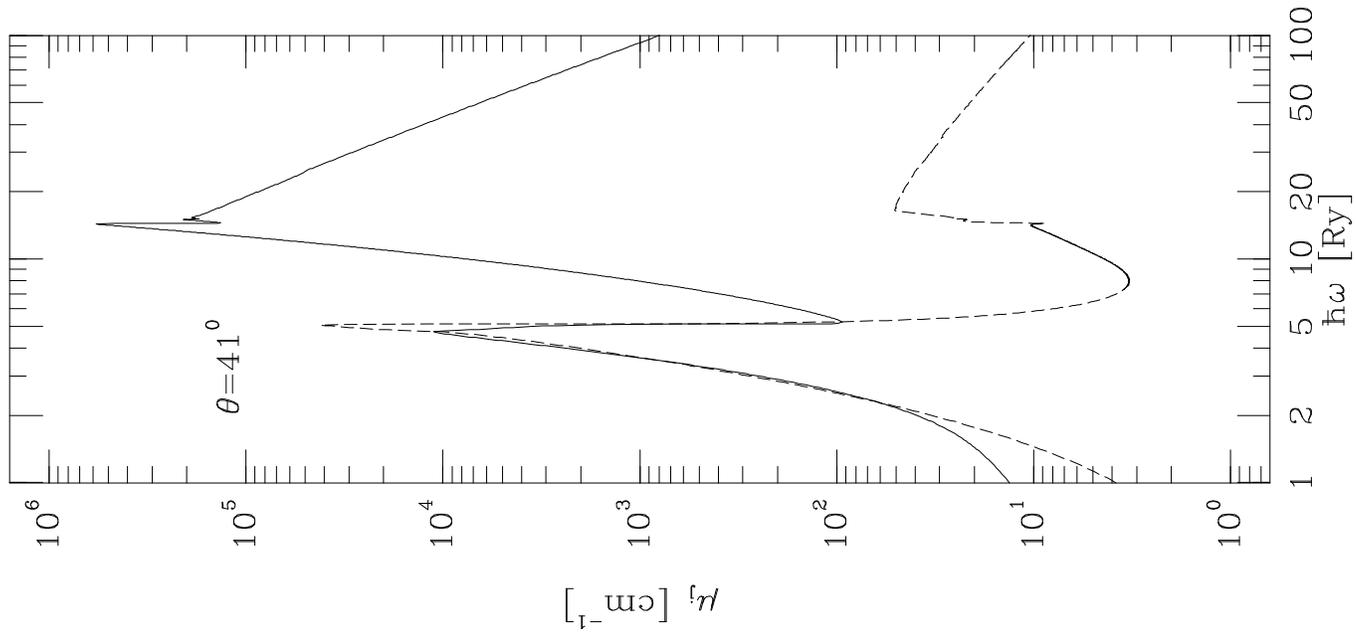
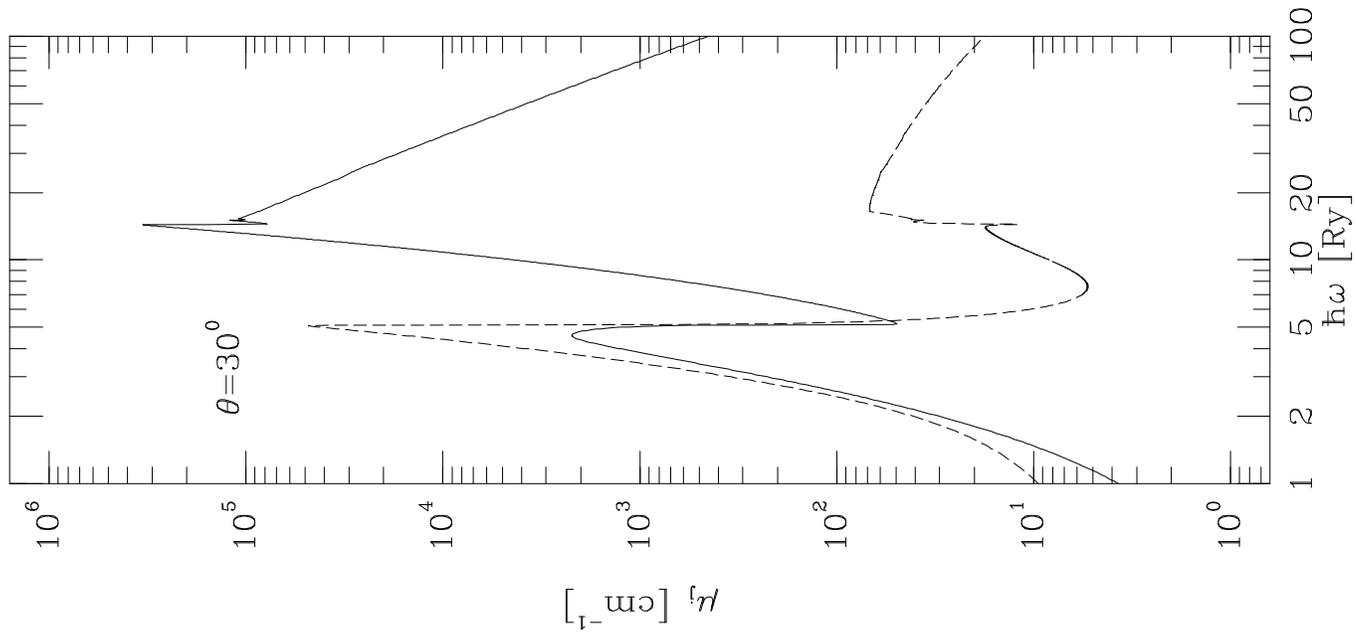
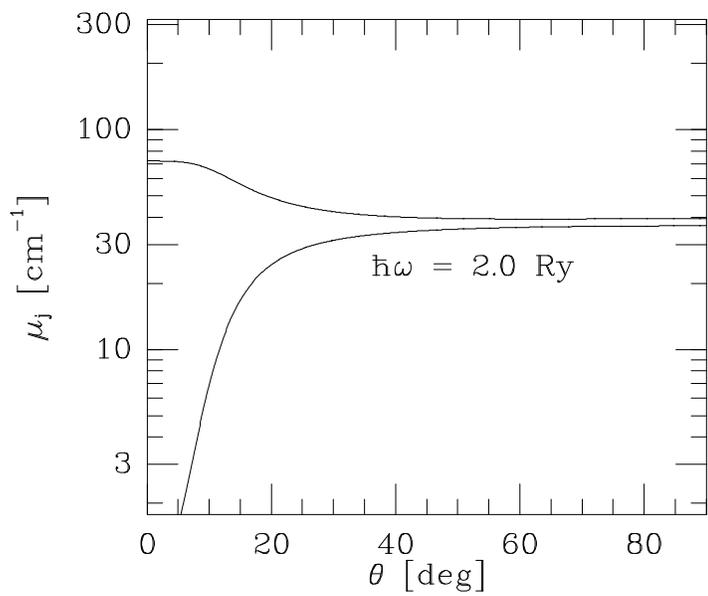
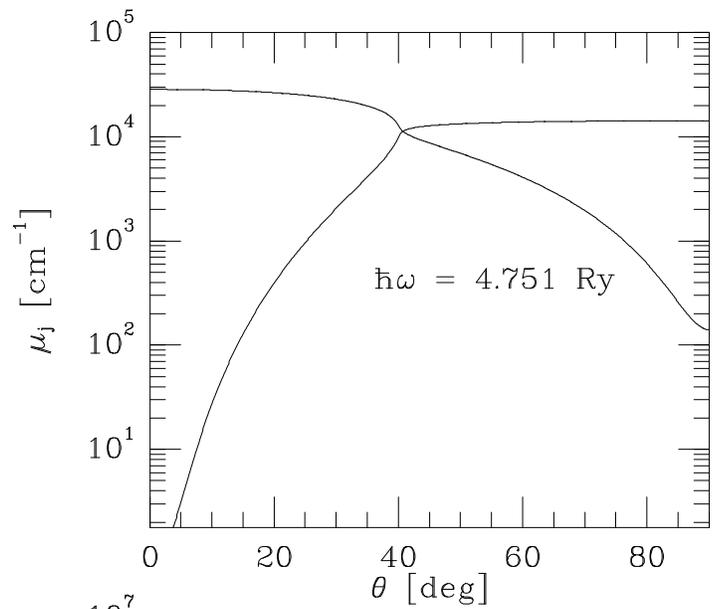
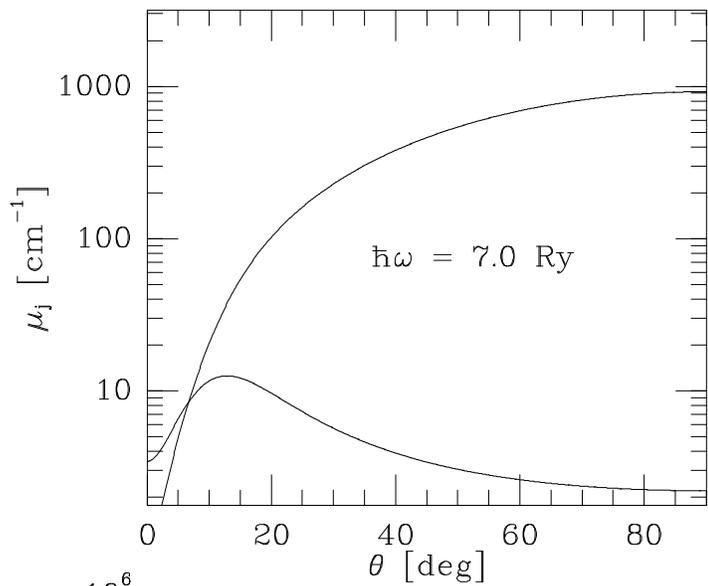
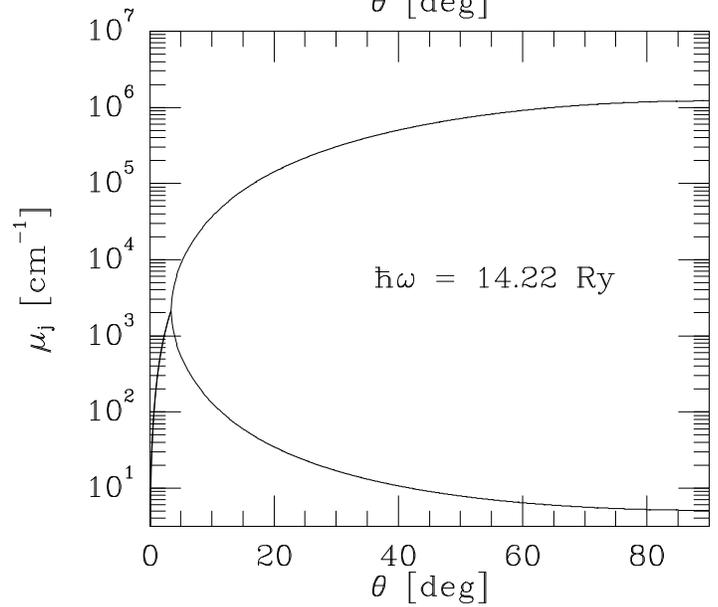
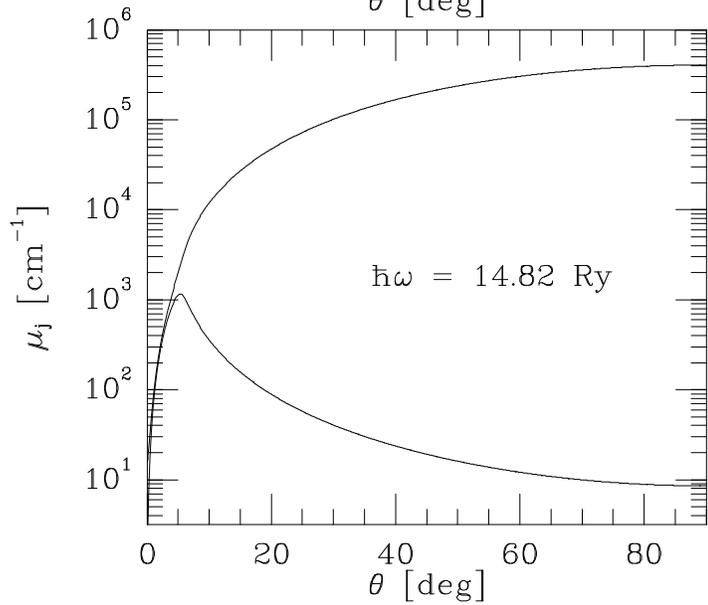
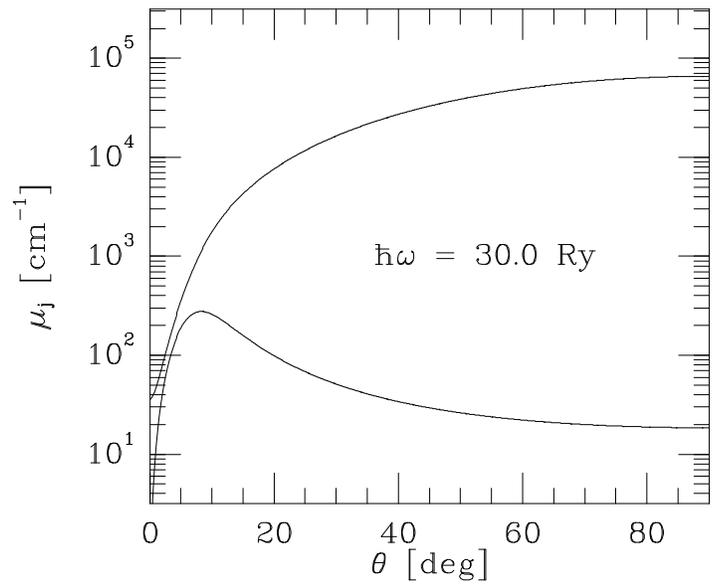
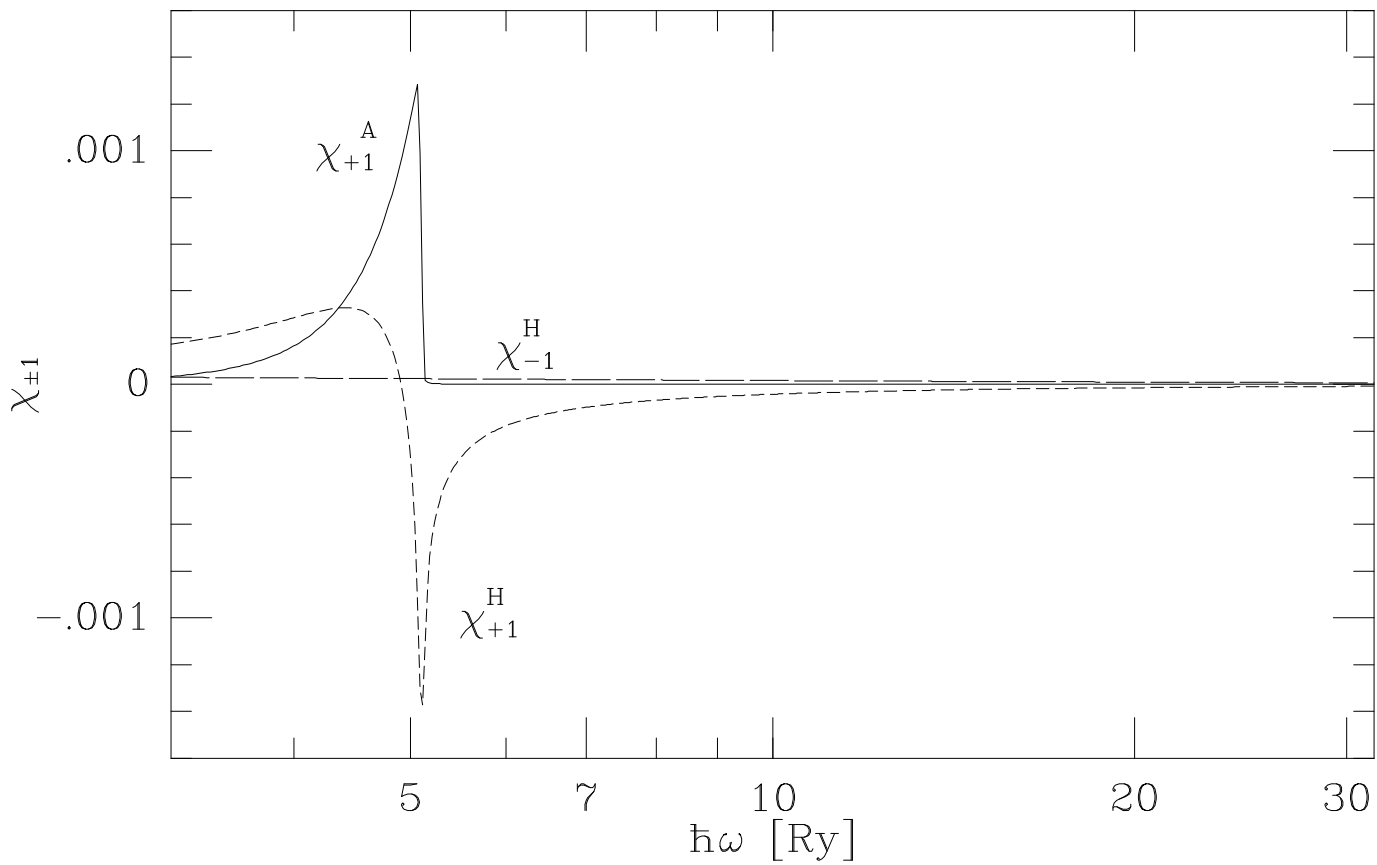
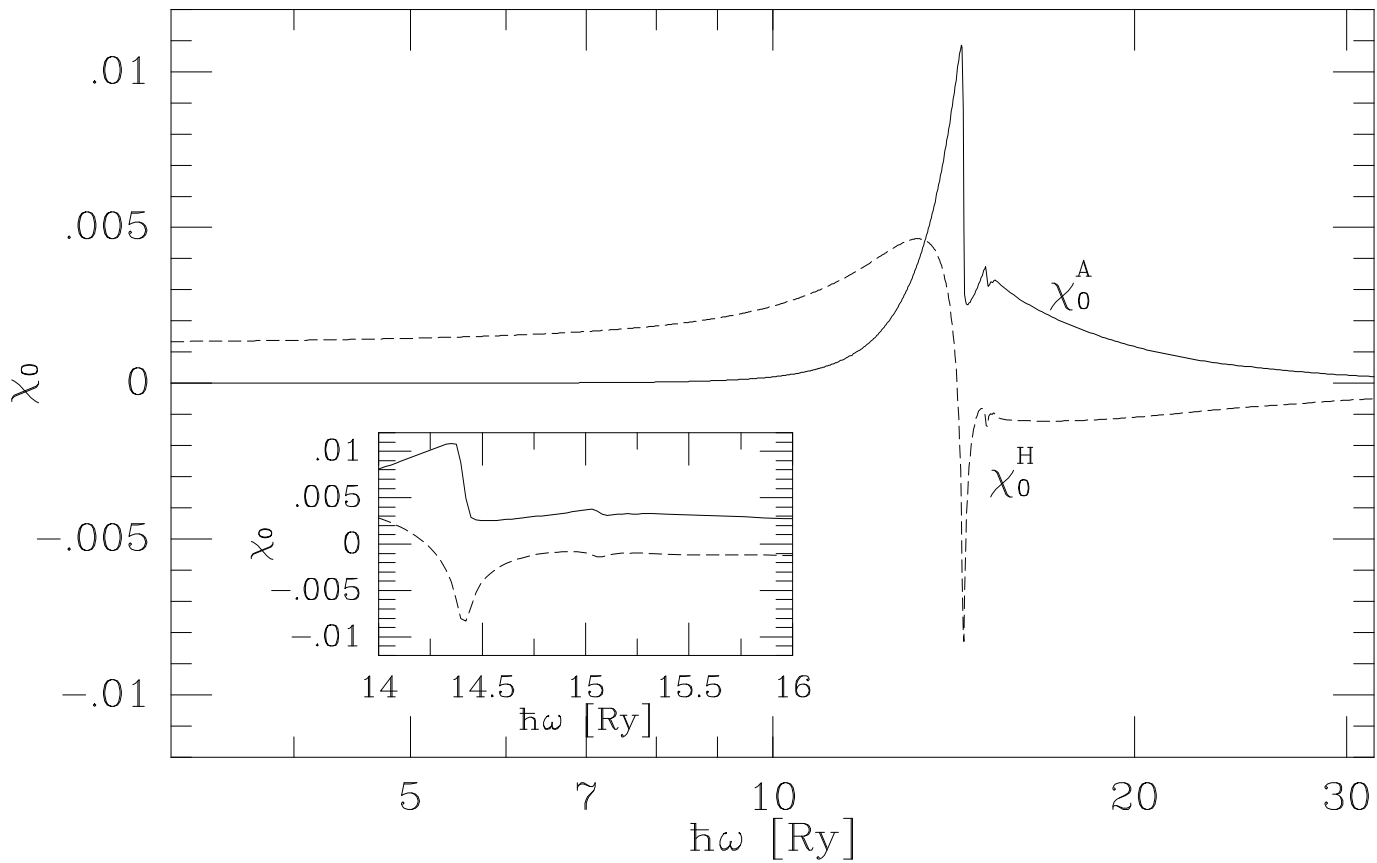
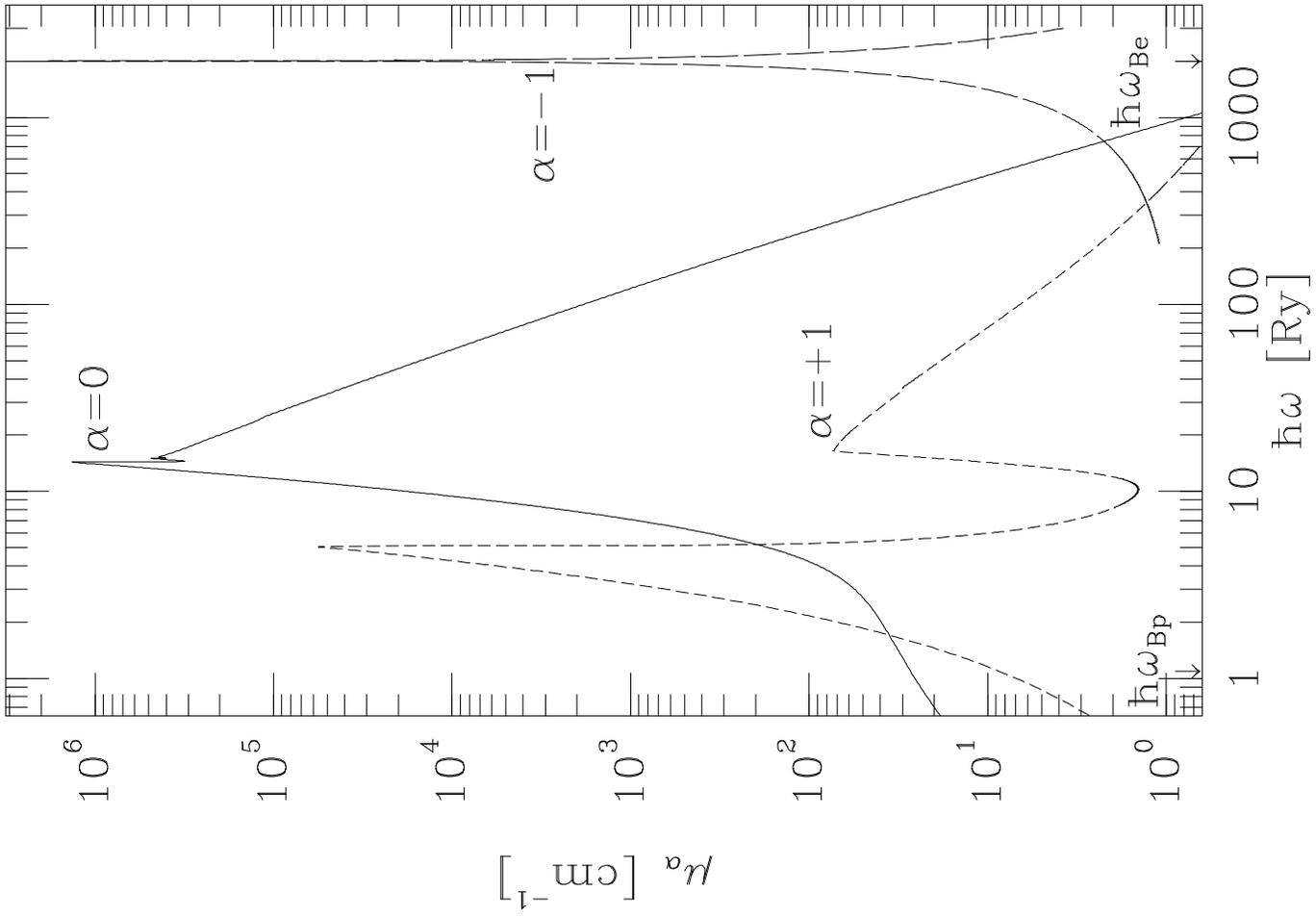
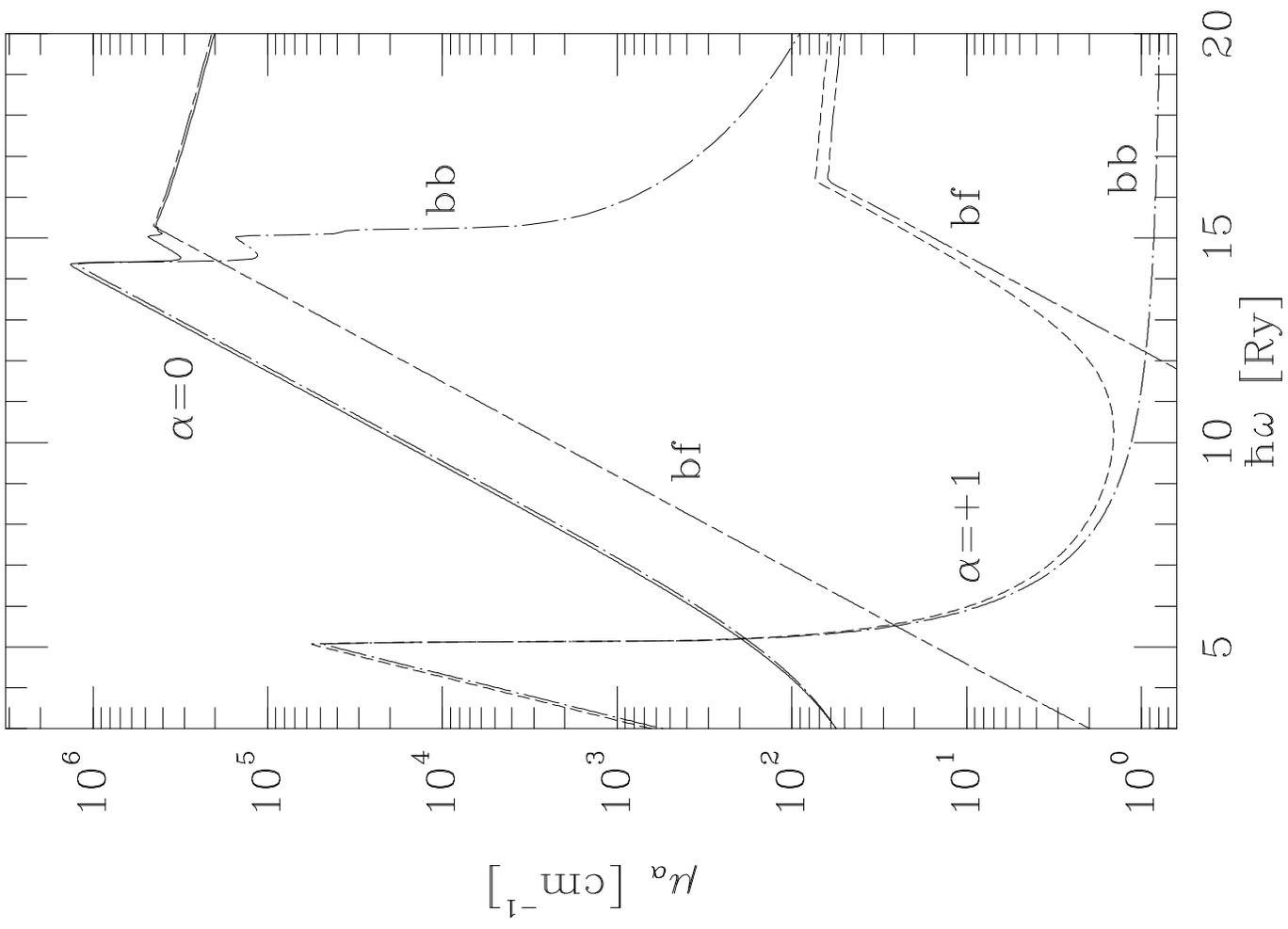
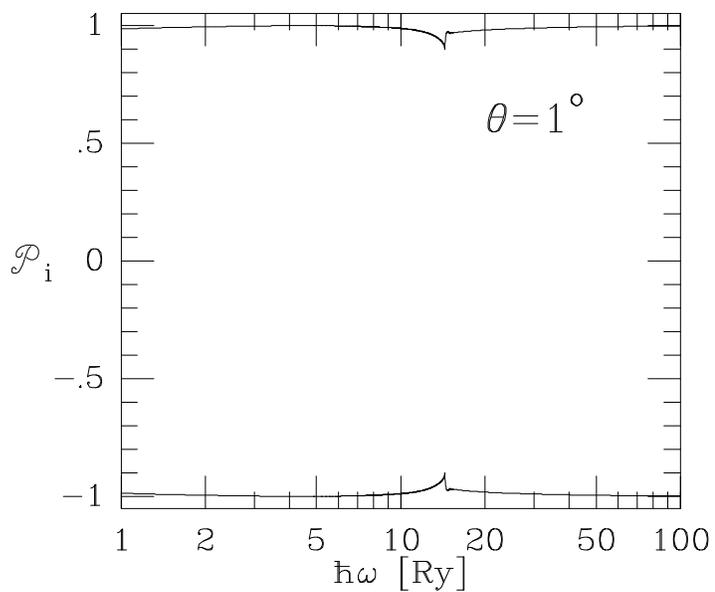
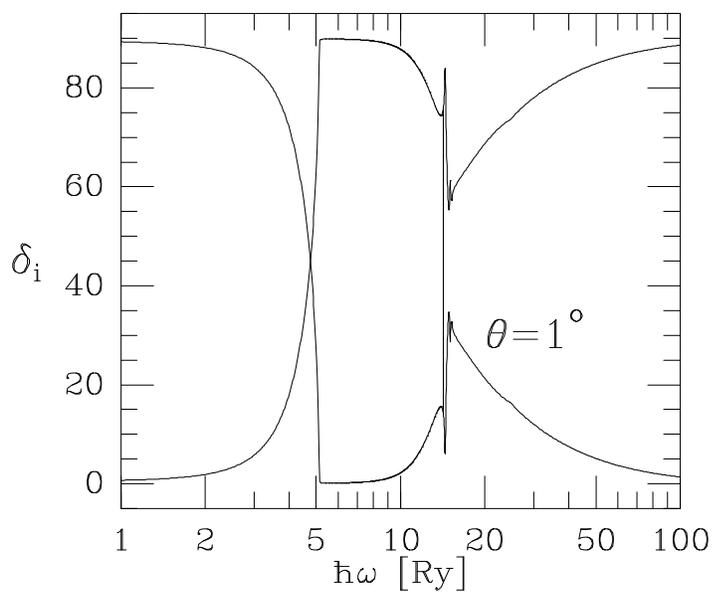
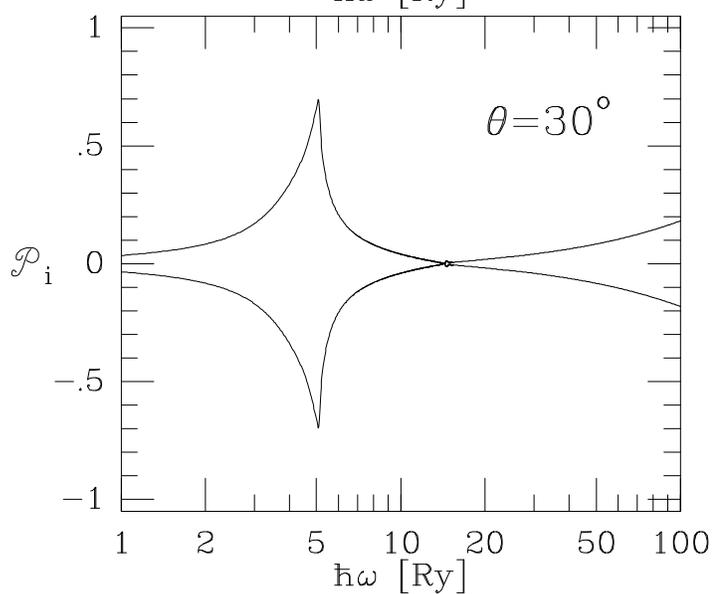
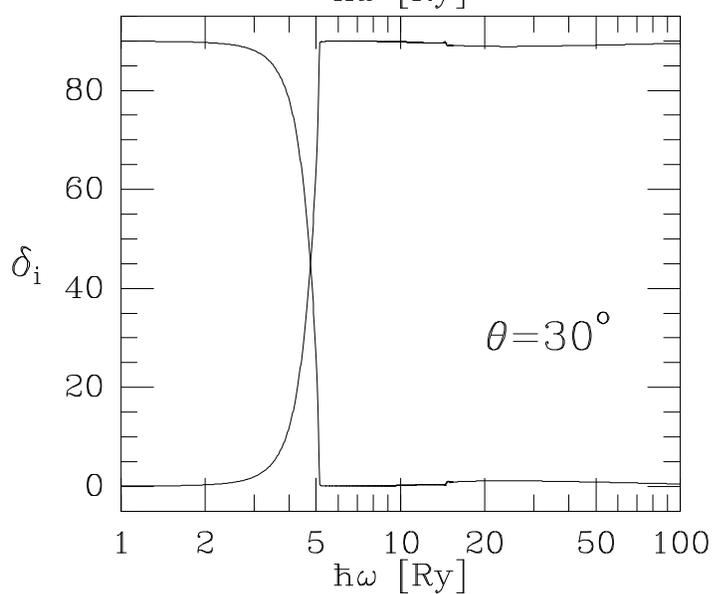
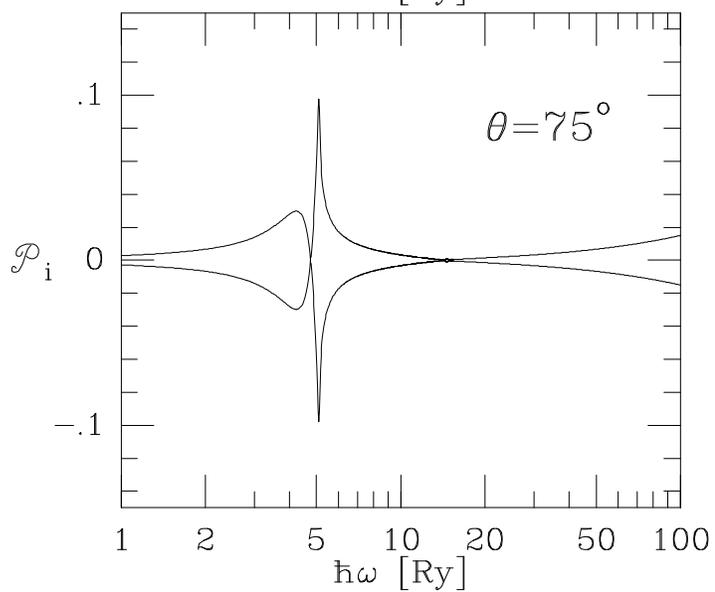
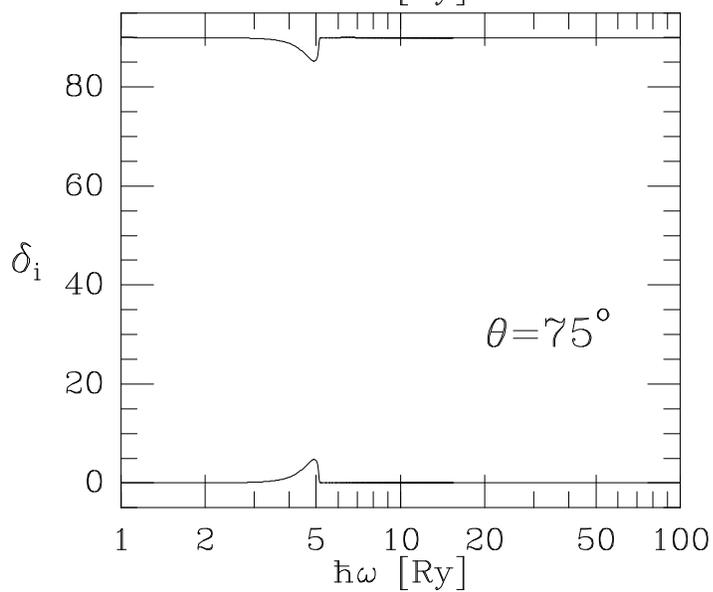
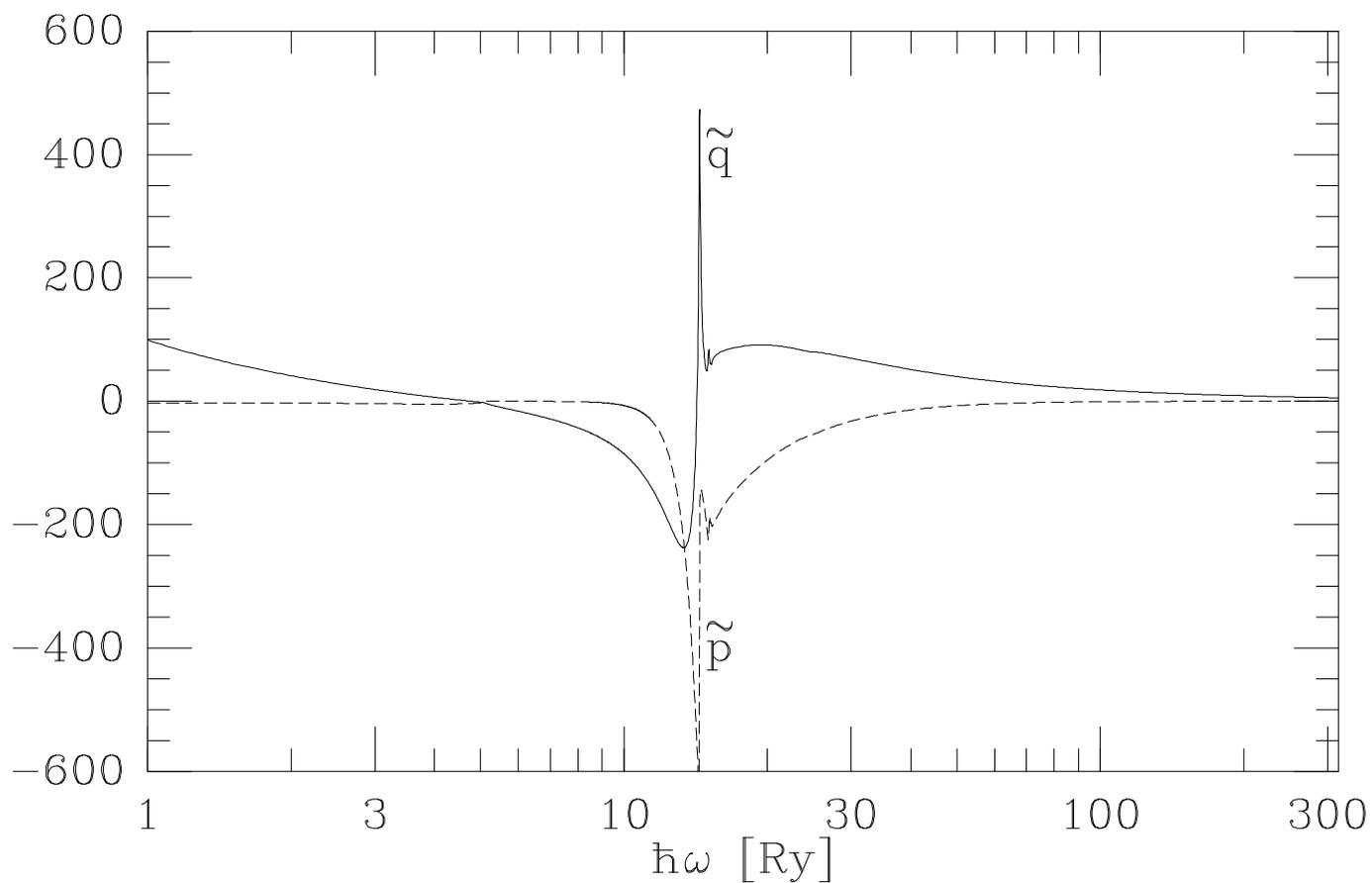
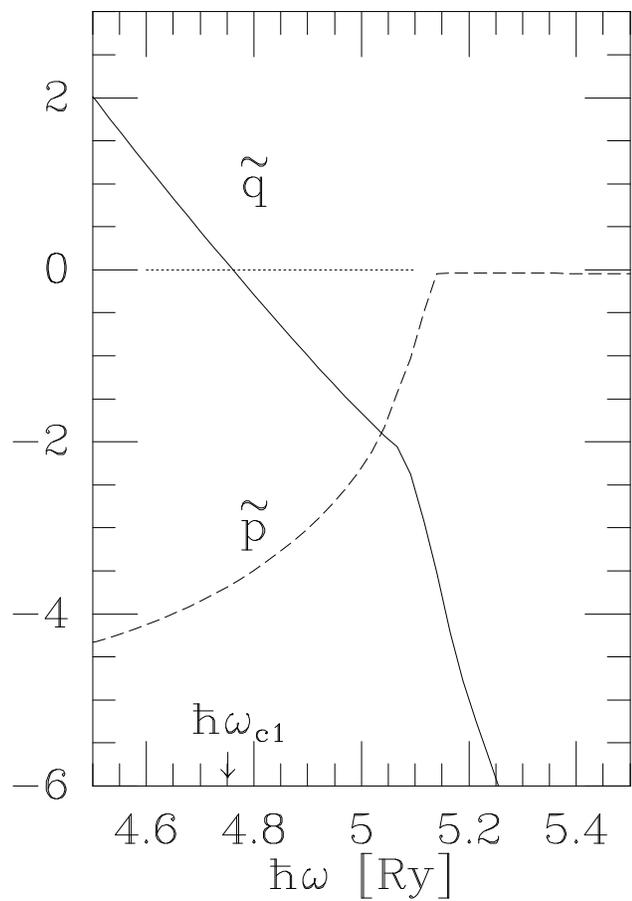
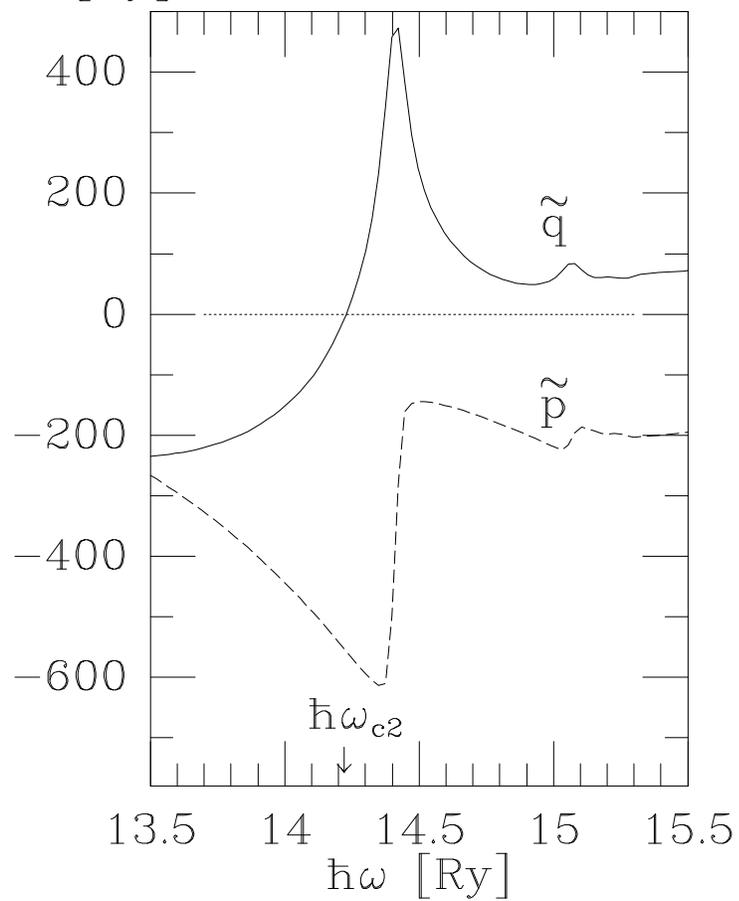
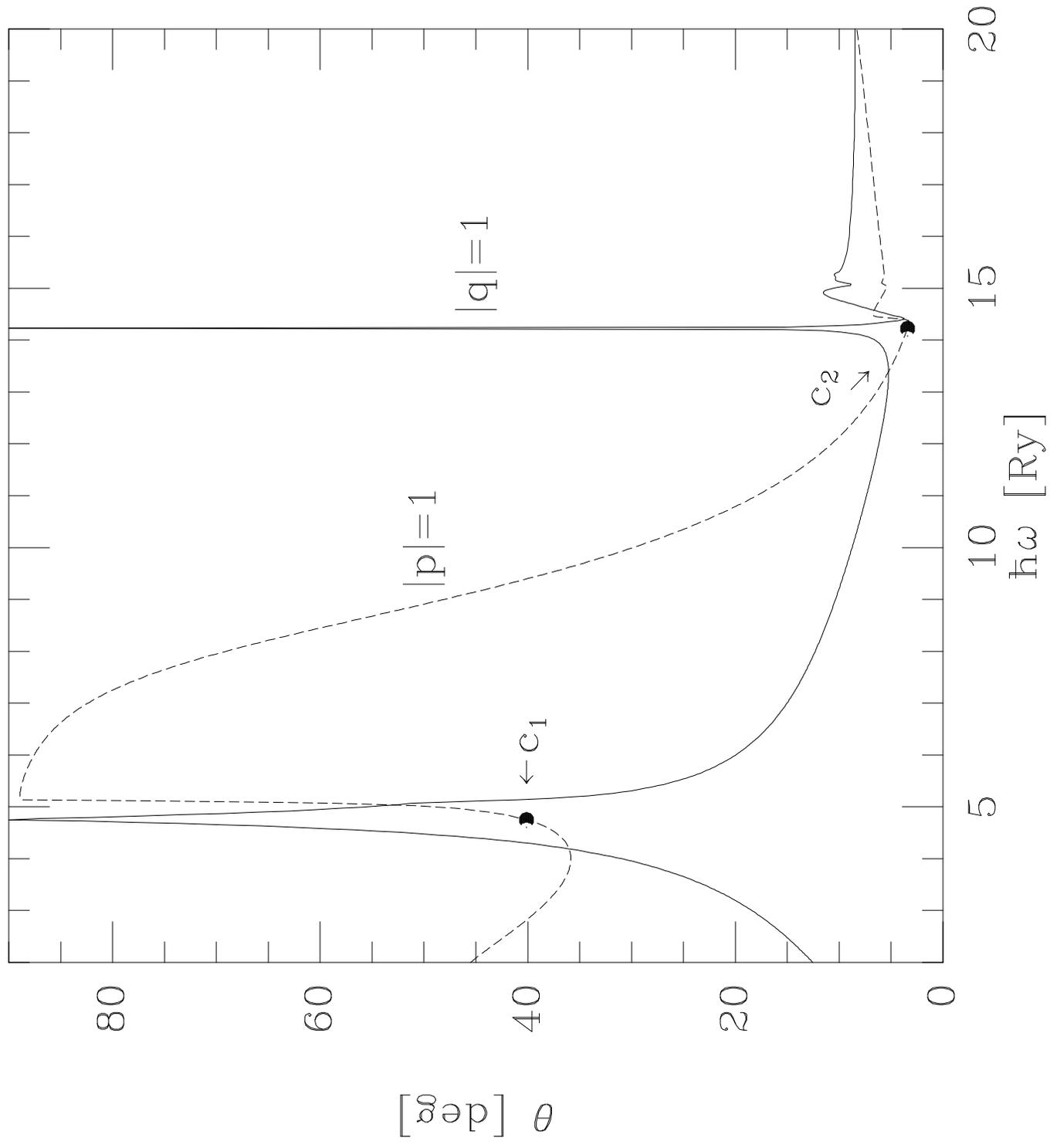
\end{document}